\documentclass[aps,pre,twocolumn,superscriptaddress,letterpaper]{revtex4}
\usepackage[pdftex]{graphicx}
\usepackage{latexsym}
\usepackage{color}
\usepackage[bookmarks,bookmarksopen,pdfstartview=FitH]{hyperref}
\hypersetup{colorlinks=true, urlcolor=blue, linkcolor=red,
  citecolor=blue}

\usepackage[sf,bf,small]{titlesec}

\usepackage{mathtools}

\usepackage{fancyhdr}
\begin{document}

\title{\bf \textsf{The Relationship of Dynamical Heterogeneity to the \\
    Adam-Gibbs and
  Random First-Order Transition Theories of Glass Formation}}

\author{\textsf{Francis W. Starr}}
\email{fstarr@wesleyan.edu}
\affiliation{Physics Department, Wesleyan University, Middletown, CT
  06459, USA}
\author{\textsf{Jack F. Douglas}}
\email{jack.douglas@nist.gov}
\affiliation{Materials Science and Engineering Division, National Institute of Standards and
Technology, Gaithersburg, Maryland 20899, USA}
\author{\textsf{Srikanth Sastry}}
\email{sastry@tifrh.res.in}
\affiliation{TIFR Centre for Interdisciplinary Sciences, Tata Institute of Fundamental Research, 21 Brundavan Colony, Narsingi, Hyderabad 500 075, INDIA
and
 Jawaharlal Nehru Centre for Advanced Scientific Research, 
Jakkur Campus, Bangalore 560064, INDIA}

\date{Submitted: October 1, 2012; ~~~Accepted: 15 January 2013}
 
\begin{abstract}
  We carefully examine common measures of dynamical heterogeneity for a
  model polymer melt and test how these scales compare with those
  hypothesized by the Adam and Gibbs (AG) and random first-order
  transition (RFOT) theories of relaxation in glass-forming liquids. To
  this end, we first analyze clusters of highly mobile particles, the
  string-like collective motion of these mobile particles, and clusters
  of relative low mobility.  We show that the time scale of the
  high-mobility clusters and strings is associated with a diffusive time
  scale, while the low-mobility particles' time scale relates to a
  structural relaxation time.  The difference of the characteristic
  times for the high- and low-mobility particles naturally explains the
  well-known decoupling of diffusion and structural relaxation time
  scales.  Despite the inherent difference of dynamics between high- and
  low-mobility particles, we find a high degree of similarity in the
  geometrical structure of these particle clusters.  In particular, we
  show that the fractal dimensions of these clusters are consistent with
  those of {swollen branched polymers or branched polymers with
    screened excluded-volume interactions, corresponding to lattice
    animals and percolation clusters, respectively.}  In contrast, the
  fractal dimension of the strings crosses over from that of
  self-avoiding walks for small strings, to simple random walks for
  longer, {more strongly interacting, strings, corresponding to
    flexible polymers with screened excluded-volume interactions}.  We
  examine the appropriateness of identifying the size scales of either
  mobile particle clusters or strings with the size of cooperatively
  rearranging regions (CRR) in the AG and RFOT theories.  We find that
  the string size appears to be the most consistent measure of CRR for
  both the AG and RFOT models.  Identifying strings or clusters with the
  ``mosaic'' length of the RFOT model relaxes the conventional
  assumption that the ``entropic droplets'' are compact.  We also
  confirm the validity of the entropy formulation of the AG theory,
  constraining the exponent values of the RFOT theory.  This constraint,
  together with the analysis of size scales, enables us to estimate the
  characteristic exponents of RFOT.
\end{abstract}

\maketitle

\section{Introduction}

One of the central mysteries of glass formation is the origin of the
dramatic increase of relaxation times approaching the glass transition
temperature, $T_g$, which is commonly interpreted as an increase of the
effective activation energy~\cite{ds-rev01, glass-rev00}.  Since the low
temperature ($T$) activation energy typically exceeds the energy of a
chemical bond, it is natural to associate this activation process with
the reorganization of multiple atoms or molecules. Indeed, there is
general agreement that glass-forming liquids are dynamically
heterogeneous, exhibiting a significant fraction of particles with
extreme high or low mobility relative to the mean, whose positions are
spatially correlated~\cite{ediger,richert,donati-pre}.  

Even before the phenomenology of dynamical heterogeneity was clearly
established, Adam and Gibbs~\cite{ag} (AG) suggested a molecular picture
of this kind in 1965, along with specific predictions for the relation
of the configurational entropy $S_{\rm conf}$ to the relaxation
dynamics. In particular, they proposed that reorganization in a liquid
occurs {\it via} hypothetical ``cooperatively rearranging regions''
(CRR), where the activation energy for relaxation is extensive in the
number of atoms or molecules that make up the CRR.  The AG model
attributes the rapid growth of relaxation time approaching $T_g$ to the
progressive growth of the CRR size on cooling. However, the AG theory
does not include a microscopic description of the CRRs, or a concrete
prescription for identifying them.  AG further argued that the
configurational entropy per CRR is roughly independent of temperature so
that the CRR mass is inversely proportional to the configurational
entropy of the fluid -- a quantity that can be estimated experimentally
by the difference of the total and vibrational entropies. Consequently,
the entropy formulation of the AG theory postulates that the
temperature-dependent activation energy for relaxation is inversely
proportional to $S_{\rm conf}$ (the ``Adam-Gibbs relationship''). This
model has proven to be highly successful to describe the $T$ dependence
of relaxation in both experiments~\cite{gt67, ra98, rclc04} (where
$S_{\rm conf}$ is estimated from specific heat measurements) and
computational studies~\cite{skt99,sastry01, speedy99, speedy01,
  hardspheres-ag, sslsss01, svsp04, mlsdst02} (where $S_{\rm conf}$ can
be formally evaluated from an energy landscape approach).

The random first-order transition (RFOT) theory~\cite{rfot89,lw07,bb04}
is based upon similar ideas to rationalize the rapid growth of
relaxation time on cooling.
In particular, the RFOT theory formulates the problem of the relaxation
in glass-forming liquids in terms of an ``entropic droplet model'', or
``mosaic'' picture, in which the liquid is divided into metastable
regions with a characteristic size $\xi$.  The balance between the
surface and bulk free energies of these regions predicts a scaling
relation between $\xi$ and $S_{\rm conf}$.  The overall barrier for
relaxation is also assumed to scale with $\xi$, providing a generalized
relationship between $S_{\rm conf}$ and relaxation.  Notably, the AG
relationship can be recovered by an appropriate limit of RFOT, so that
these models are potentially directly linked.  As in the case of the AG
theory, RFOT theory does not provide a specific molecular definition of
the length scale of collective motion.  Thus, both the AG and RFOT
approaches leave the precise nature of cooperative rearrangements and
their relation to dynamical heterogeneity open to interpretation and
quantification.

Computer simulations have been particularly helpful to quantify the
nature of dynamical heterogeneity, as the spatial and temporal
heterogeneity of glass-forming fluids is difficult to probe directly
with experiments.  It is now appreciated that atoms or molecules of
extreme mobility (or immobility) tend to cluster, and that the most
mobile clusters can be further divided into groups of atoms or molecules
that move cooperatively in a roughly co-linear, or string-like
fashion~\cite{strings-blj, strings-polymer, gvg04, rigg06,
  vdhg04,ssds00, gssbs02}, and this phenomenon has been confirmed
experimentally in colloidal particle tracking measurements.~\cite{msr99,
  exp-str2D1, exp-str2D2, exp-str3D}.  Consistent with the ideas of the
AG and RFOT theories, the sizes of clusters and strings grow on cooling
toward $T_g$, but it is not clear if either of these structures are
appropriate measures of the size scales envisioned by these theories.
Earlier works have considered both the possibilities of using the mobile
particle clusters or the strings as the CRR of AG, and each of these
studies indicated promising results~\cite{gbss03,gvbmsg05,sd11,pds13}.
We should point out that there are other ways to characterize the length
scales of heterogeneity.  In particular, the use of a four-point
correlation function offers an approach rooted in the framework of
statistical mechanics that reveals a growing dynamical length scale on
cooling~\cite{kt88, dasgupta91, fdpg99, lssg03, berthier05, bbmr06,
  kds09, kds10, szamel10}.

In this work, we systematically dissect cluster and string-like nature
of the heterogenous motion in a model glass-forming polymer melt, and
then consider what measure or measures of dynamical heterogeneity, if
any, may appropriately quantify the size scales envisioned by the AG or
RFOT approaches. In doing so, we expand on a general methodology to
identify subsets of extreme immobility.  Our results span a broad
temperature range, from very high $T$, to somewhat below the crossover
temperature $T_c$ often associated with mode-coupling theory.  We find
that, at the characteristic time of maximal clustering, the structures
of mobile and immobile clusters exhibit statistical properties that are
consistent with the properties of equilibrium branched polymers, which
are the same as clusters approaching a percolation transition (lattice
animals).  When mobile clusters are decomposed into strings, the
geometry of short strings are consistent with self-avoiding walks, while
larger strings (that appear at low temperature) behave like simple
random walks.  We find that none of the cluster types that we study form
compact objects when examined at their respective characteristic times.
Moreover, the characteristic times of mobile and immobile clusters
provide a physically transparent way to understand decoupling phenomena,
as the mobile cluster time scales have essentially the same temperature
dependence as diffusive time scales, while the immobile cluster time
scale follows the structural relaxation time.  We consider both mobile
clusters and strings as possible descriptions of CRR in the AG and RFOT
models, and find that the strings -- which necessarily incorporate large
mobility and cooperativity of displacement -- best accord with the
quantitative description of the mass or length scales of cooperative
clusters described by both these theories.

\section{Model and Simulation Details}

Our results are primarily based on molecular dynamics simulations of a
melt containing 400 chains of ``bead-spring'' polymers, each chain
consisting of 20 monomers~\cite{fene1}.  At this length, the chains are
unentangled.  All monomers interact via a force-shifted Lennard Jones
(LJ) potential, truncated at 2.5$\sigma$ so that dispersive attractions
are included ($\sigma$ is the LJ length parameter).  Neighboring
monomers along a chain also interact via a FENE spring potential to
create covalent bonds.  The FENE parameters are $k=30\epsilon$ and $R_0
= 1.5\sigma$, chosen to create a mismatch in the length scale of bonded
and non-bonded interactions, thus frustrating crystallization and making
the model a good glass former~\cite{basch-soft-rev10}.  All values are
reported in reduced LJ units.  Standard units for temperature are
recovered by multiplying $T$ by $\epsilon/k_B$, where $k_B$ is
Boltzmann's constant.  Time is given in units of $(m \sigma^2 /
\epsilon)^{1/2}$.  The simulations cover the range of $0.3<T<2.5$ at
constant density $\rho = 1.0$.  For all $T<1$, we perform five
independent simulations to improve statistics.  Each simulation consists
of an equilibration run followed by data collection; the duration of
each run is determined from the $\alpha$ relaxation time (discussed
below) to ensure we sample only equilibrium states.  Temperature is
controlled via the Nose-Hoover algorithm, which is implemented via the
rRESPA method using a time step of 0.002 for bond forces with 3 updates
for each non-bonded force update~\cite{frenkel-book}.

\begin{figure}[t]
\includegraphics[clip,width=3.2in]{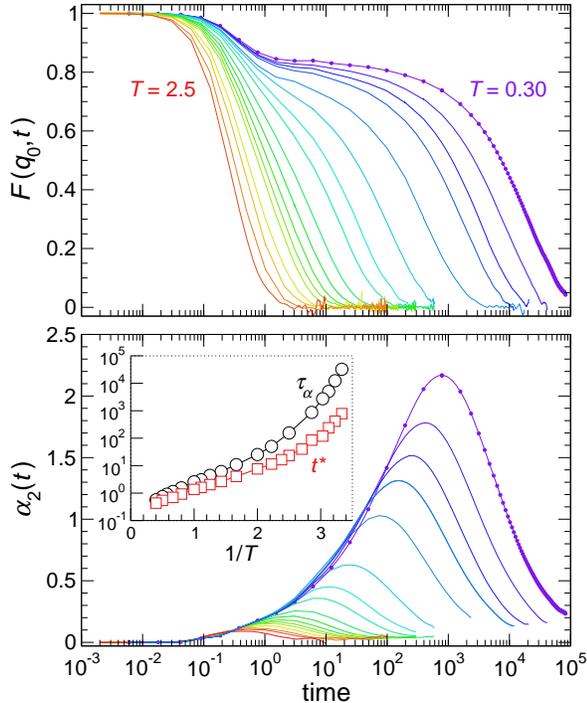}
\caption{Characterization of basic dynamical properties of the polymer
  melt. (a) The coherent density-density correlation function $F(q_0,t)$
  for all $T$.  The $\alpha$-relaxation time $\tau_\alpha$ is defined by
  $F(q_0,\tau_\alpha) = 0.2$.  Symbols are shown for the lowest $T$ to
  indicate typical intervals at which data are collected. (b) The
  non-Gaussian parameter $\alpha_2(t)$ at each $T$ shows a peak due to
  the correlated motion that occurs roughly on the time scale of $t^*$,
  defined by the maximum of $\alpha_2(t)$. The inset compares the
  behavior of $\tau_\alpha$ and $t^*$. }
\label{fig:fqt+alpha2}
\end{figure}

The dynamics of this model (or the closely related model that excludes
LJ attractions) have been extensively studied in previous
simulations~\cite{basch-soft-rev10}.  To provide basic characterization
for subsequent detailed analysis, we first consider the relaxation of
the coherent density-density correlation function $F(q,t)$
(Fig.~\ref{fig:fqt+alpha2}(a)).  We evaluate $F(q,t)$ at the wave vector
$q_0$ corresponding to the first peak of the structure factor where
relaxation is slowest (except for the limit $q\rightarrow 0$).
We define the $\alpha$-relaxation time by $F(q_0,\tau_\alpha) =
0.2$. The $T$-dependence of $\tau_\alpha$ (Fig.~\ref{fig:fqt+alpha2}
inset) is characterized by simple Arrhenius behavior for $T>T_A$; for
$T_g<T<T_A$, $\tau_\alpha$ grows significantly faster on cooling, and is
well-approximated by the ubiquitous Vogel-Fulcher-Tamman (or
Williams-Landau-Ferry) expression
\begin{equation}
\tau_\alpha = \tau_0 \exp\left[\frac{DT_0}{T-T_0}\right]
\label{eq:vft}
\end{equation}
where $T_0$ is an extrapolated divergence temperature that is typically
slightly below the laboratory glass transition temperature $T_g$, and
$D$ characterizes the curvature (or fragility) of $\tau_\alpha$.  From
our simulations, the crossover from Arrhenius behavior $T_A \approx 0.8$
and $T_0 = 0.20 \pm 0.01$.  For this system and density, the
characteristic temperature $T_c$ associated with power-law behavior
$\tau \sim (T-T_c)^{-\gamma}$ has been estimated to be $T_c
=0.35$~\cite{ssdg02}; thus we probe $T$ significantly below $T_c$.
Additionally, we also know that $T_g \approx 1.2\, T_0$ from the simple
and widely used convention $\tau_\alpha(T_g) = 100$~s~\cite{sd11}.  For
a simple polymer (like polystyrene) with $T_g \approx 100~^\circ$C, the
reduced units can be mapped to physical units relevant to real polymer
materials, where the size of a chain segments $\sigma$ is typically
about 1~nm to 2~nm, time is measured in ps, and $\epsilon \approx
1$~kJ/mol.

Since we will examine in detail the spatial heterogeneity of the
segmental dynamics, we also evaluate the non-Gaussian parameter
$\alpha_2$ as a basic indicator of the time scale and strength of
correlated motion (Fig.~\ref{fig:fqt+alpha2}(b)).  The peak of
$\alpha_2$ defines the time $t^*$, which provides a characteristic time
scale of the spatially heterogeneous motion.  The amplitude of the peak
of $\alpha_2$ also increases, a consequence of the increasing degree of
spatial correlations of the motion.  {Although it is not explicitly
  documented, it is implicit from many previous works~\cite{ka95-1,
    sgtc96, kdppg97, donati-pre, gssg01} that $t^*$ grows less rapidly
  than $\tau_\alpha$ on cooling, as confirmed in the inset of
  fig.~\ref{fig:fqt+alpha2}(b)}. In other words, these characteristic
times ``decouple''.  {This can be expected since $t^*$ is a
  diffusive time scale (see appendix~\ref{app:non-gaussian}), and the
  diffusion coefficient $D$ has long been known to decouple from
  structural relaxation.~\cite{ediger}} AG never envisioned that
glass-forming liquids should be characterized by multiple relaxation
times; consequently, they did not distinguish between mass diffusion and
momentum diffusion ({\it i.e.,}\ viscous relaxation), but their language
clearly relates to modeling mass diffusion.  Fortunately, since these
time scales maintain a fractional power-law relation over a large range
extending from $T_g$ to $T_A$, the AG (or RFOT) approaches can be
equally applied to either mass or momentum diffusion, a point that we
expand upon below.

\section{Dynamical Clusters Approaching the Glass Transition}

It is widely appreciated that, below the onset temperature $T_A$, the
dynamics become increasingly spatially heterogeneous approaching $T_g$.
Regions with either enhanced mobility or diminished mobility form in a
spatially correlated manner, and the motions within mobile regions can
be further dissected into more elementary groups that move in a
string-like, cooperative fashion.  In this section, we examine several
ways to characterize correlations in mobility and analyze the geometry
of these structures.

\subsection{Mobile and Immobile Clusters}

Since the distribution of particle mobilities varies continuously, the
first challenge is how to distinguish mobility subsets.  For a variety
of systems~\cite{tm93, kdppg97,donati-pre,gssg01}, it has been
shown that choosing the subset of particles that that have moved farther
than is expected from the Gaussian approximation at the characteristic
time $t^*$ offers a useful metric to identify the highly-mobile
particles.  Depending on the system, these mobile particles typically
account for 5~\% to 7 \% of the particles below $T_A$.  This is also
true for the present system, confirmed in our own calculations and in
ref.~\cite{gssg01}.  Accordingly, we follow the choice of
ref.~\cite{gssg01} where the same model was examined, and select mobile
particles as the 6.5~\% of particles with the greatest displacement over
any chosen interval $t$.  This allows us so see the evolution of mobile
particle properties over all $t$, in addition to the characteristic time
$t^*$.

At the other mobility extreme, we identify particles of extreme {\it
  immobility}.  While a variety of methods to identify low mobility
particles have been explored in past literature~\cite{donati-pre, vz05,
  vkbz02, dsz02, tanaka05, tanaka10, weitz06, am07, psdh10}, there is
considerable variation in the details of these approaches.  Moreover,
many studies of immobile particles do not condition the selection on
mobility, but rather on local packing considerations ({\it e.g.},
icosahedral packing, Frank-Kasper clusters, a sufficient number of
neighbors {\it etc.}). Such attempts are potentially valuable for
relating structure to dynamical behavior, but in the present work we
wish focus purely on dynamical considerations that should be applicable
to characterizing mobility all glass-forming liquids, rather than any
particular fluid having its own unique type of local ordering.
Consequently, we have developed a criterion for immobile particles based
on the tendency for ``caged particles'' to cluster.  We provide a
detailed description of the method in appendix~\ref{appendix:immobile}
to avoid breaking the flow of our main results.  Broadly speaking, we
can identify caged particles for any time $t$ by those particles with
displacements smaller than the (weakly $T$-dependent) plateau value
observed in the mean-squared displacement.  We then find the time at
which these caged particles form the largest clusters and evaluate what
fraction of the system these caged particles constitute.  Similar to the
approach for mobile-particle clusters, we fix this fraction for all $t$
to track the evolution of the clustering of the immobile particles.
Note that the fraction of immobile particles from this method is
$T$-dependent, increasing from $\approx$~5~\% at the lowest $T$ studied
up to 11~\% at $T_A$.  Above $T_A$, the cage size is not well defined.

\begin{figure}[tb]
\includegraphics[clip,width=3.2in]{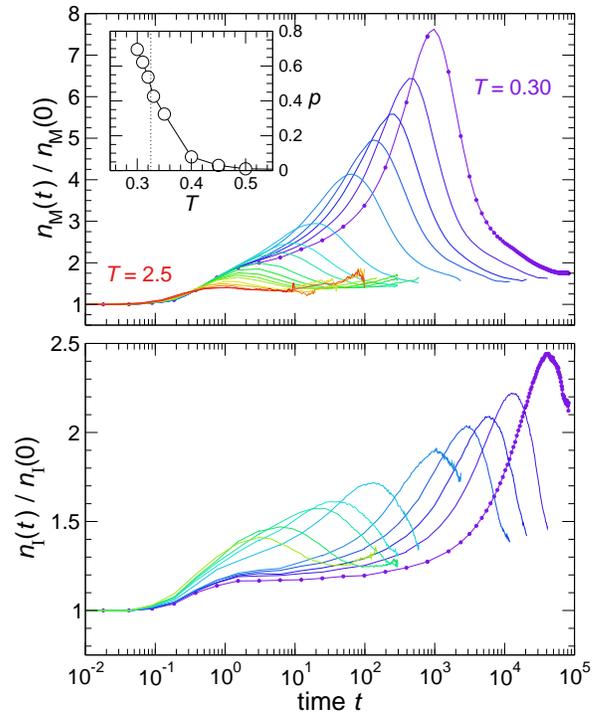}
\caption{The dynamical average cluster size for (a) high-mobility
  particles $\langle n_M(t)\rangle$ and (b) low-mobility particles
  $\langle n_I(t) \rangle$ at all $T$ studied. The data are normalized
  by the value at $t=0$, equivalent the cluster size of the same
  fraction of particles chosen at random.  The definitions of the
  mobility groups are discussed in the main text.  The inset of part (a)
  shows the percolation probability $p$ of mobile particle clusters as a
  function of $T$; the dotted vertical line indicates the temperature
  where $p\approx 0.5$, a standard identifier of the percolation
  transition in finite-sized systems~\cite{stauffer}.  For immobile
  particle clusters, $p<0.2$ for all $T$, so that percolation is not
  prevalent. }
\label{fig:all-clusters}
\end{figure}

Having identified the most and least mobile particles at each interval
$t$, we examine the average cluster size of the mobile $\langle
n_M(t)\rangle$ and immobile $\langle n_I(t)\rangle$ subsets.  We plot
the cluster sizes (fig.~\ref{fig:all-clusters}) relative to the cluster
size of the same fraction of particles chosen randomly; for immobile
particle clusters, this eliminates the trivial $T$-dependence of
immobile particle cluster size that arises from $T$-dependence of the
fraction of immobile particles (see the appendix~\ref{appendix:immobile}
for further discussion).  We define a cluster by the group of
nearest-neighbor particles that have a separation less than the
nearest-neighbor distance, given by the distance of the first minimum
$r_{\rm min} = 1.46$ in the density-density pair correlation function.
Fig.~\ref{fig:all-clusters} shows the typical behavior for mobile
particle clusters; namely, $\langle n_M(t)\rangle$ peaks at a
characteristic time $t_M$ that increases on cooling, and that the peak
value $\langle n_M(t_M)\rangle$ also grows on cooling, indicating an
increase in the spatial extent of correlations.  The immobile particle
clusters exhibit the same qualitative trend.  Note that we plot the
average cluster size, not the {\it weight}-averaged cluster size; the
qualitative behavior of both are the same.

There are significant differences between $\langle n_M(t)\rangle$ and
$\langle n_I(t)\rangle$ to consider. First, we see that the relative
peak size of the mobile particle clusters is larger and increases more
rapidly on cooling than that of the least mobile clusters. At the lowest
$T$ studied, the mobile particle clusters become so large that the
percolation probability $p$, defined as the fraction of configurations
with a spanning cluster, approaches unity (inset
fig.~\ref{fig:all-clusters}(a)). If we define the percolation threshold
by $p_c = 0.5$ (as is common in finite systems~\cite{stauffer}), the
percolation temperature $T_p\approx 0.32$ for mobile particle
clusters. Consequently, we likely underestimate the size of mobile
particle clusters at the three lowest $T$ studied. For the immobile
particles clusters, $p<0.2$ for all $T$, so that finite-size effects
should not be of concern.

\begin{figure}[t]
\includegraphics[clip,width=3.2in]{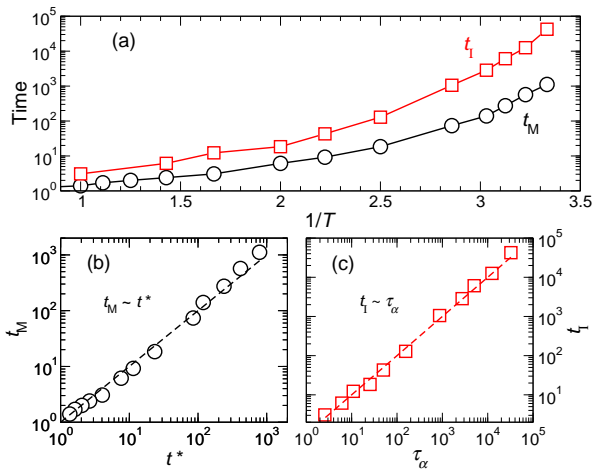}
\caption{(a) Characteristic time where the cluster size peaks for mobile
  (black circles) and immobile (red squares) clusters. Note the
  similarity to the behaviors of $\tau_\alpha$ and $t^*$, shown in
  fig.~\ref{fig:fqt+alpha2}(b).  Panels (b) and (c) are parametric plots
  that show that $t_M \sim t^*$ and $t_I \sim \tau_\alpha$; the dashed
  lines indicate an equality between these quantities.  
 }
\label{fig:cluster-timescales}
\end{figure}

The characteristic time scales of these cluster types differ
significantly at low $T$.  Specifically, the time scale for the peak of
the mobile particle clusters $t_M$ is significantly smaller than the
peak time $t_I$ of the immobile particle clusters
(fig.~\ref{fig:cluster-timescales}), similar to the difference in the
time scale between $t^*$ and $\tau_\alpha$.  Indeed, parametrically
plotting these quantities shows that $t_M \sim t^*$ and $t_I \sim
\tau_\alpha$ (fig.~\ref{fig:cluster-timescales}(b) and (c)), and in fact
the respective quantities are nearly equal. The similarity between $t_M$
and $t^*$ has been previously noted~\cite{gvbmsg05}. The linear scalings
can be understood in the context of decoupling phenomena. Specifically,
it is widely observed that $\tau_\alpha$ grows more rapidly on cooling
than the time scale associated with the diffusion coefficient $D$,
giving rise to a breakdown in the Stokes-Einstein relation $D/T \sim
\tau_\alpha$ -- the same decoupling phenomenon previously discussed for
$t^*$. Decoupling can be qualitatively understood as a consequence of
dynamical heterogeneity, since $D$ will be dominated by the most mobile
particles, while relaxation functions (and hence $\tau_\alpha$) will
arise from the least mobile particles. Consequently, we anticipate, and
indeed observe, decoupling between the $t_I$ and $t_M$ timescale that
matches the decoupling between the $t^*$ and $\tau_\alpha$ timescales.

\subsection{Cluster Size Distribution and Fractal Dimension}

\begin{figure*}[tb]
\includegraphics[clip,width=2.2in]{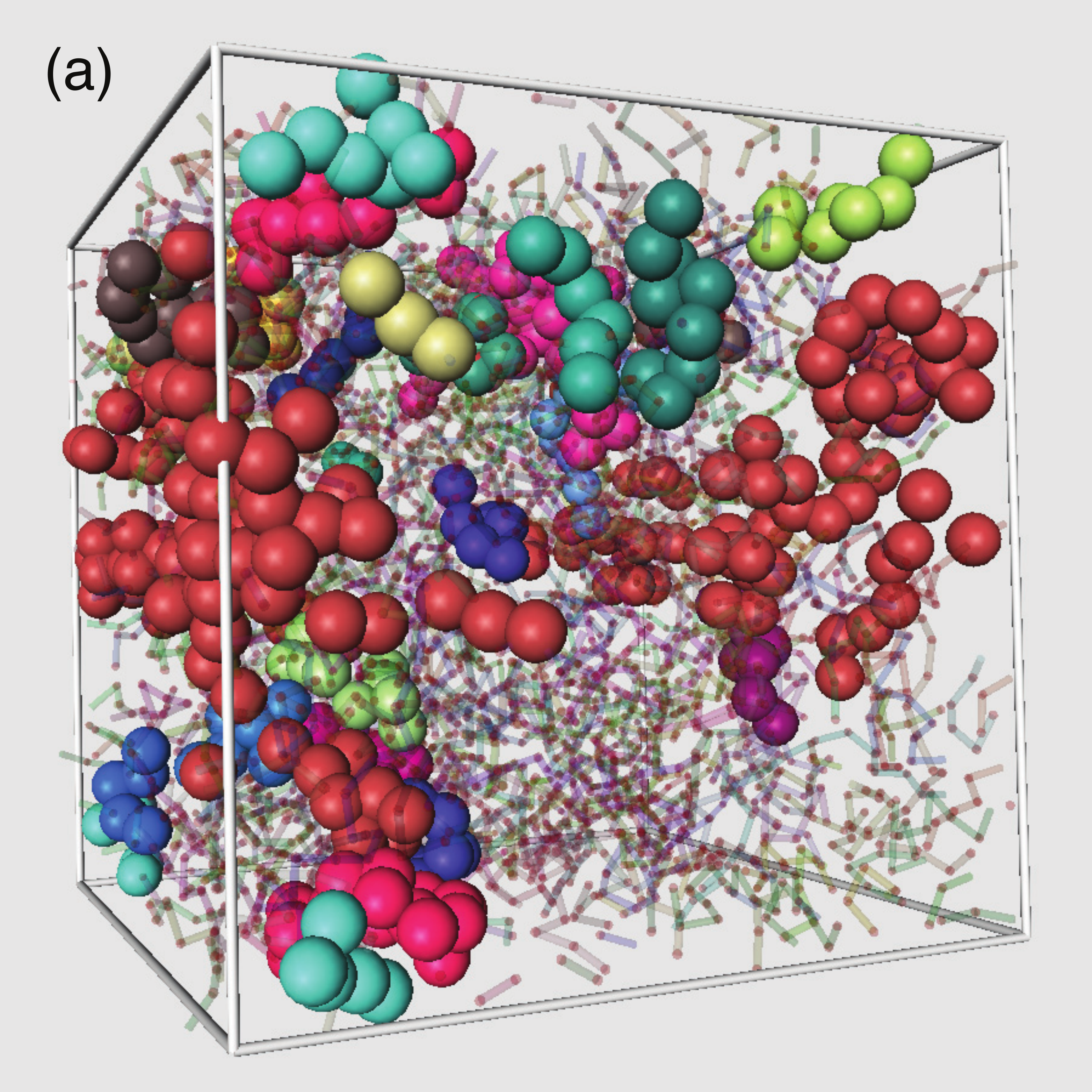}
\includegraphics[clip,width=2.2in]{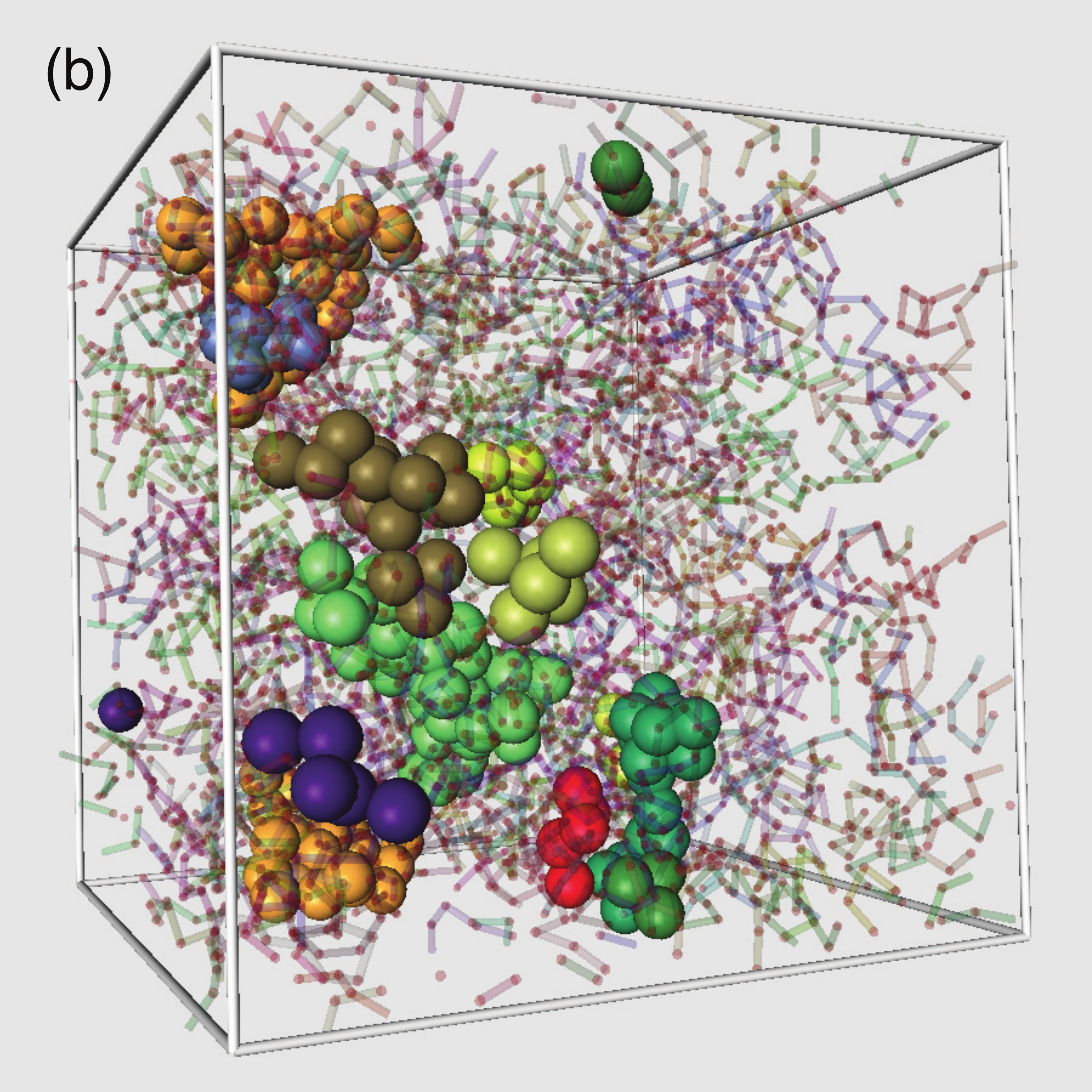}
\includegraphics[clip,width=2.2in]{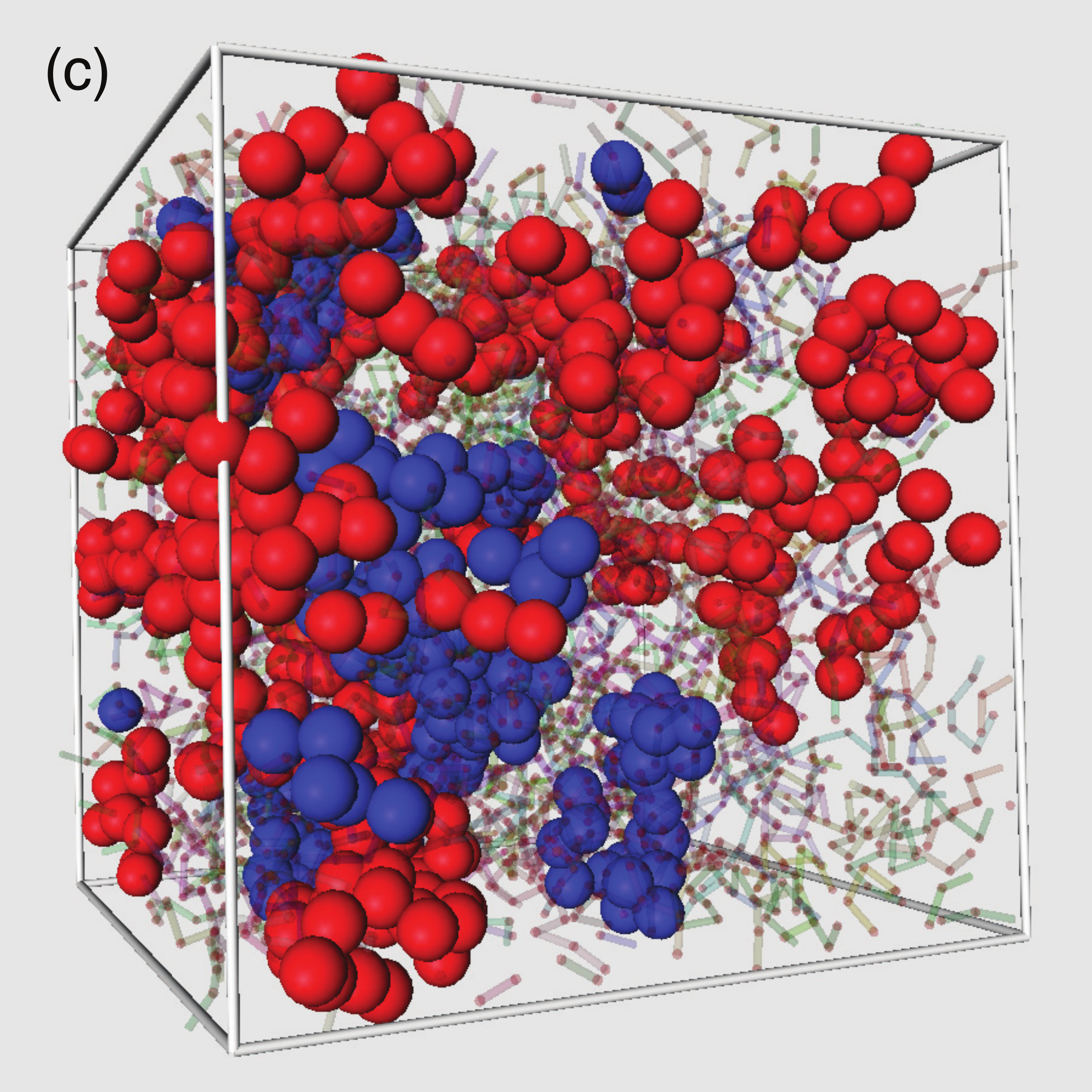}
\caption{Typical examples of (a) the most mobile and (b) least mobile
  clusters.  Different clusters are shown in different colors, and the
  segments of all chains are shown translucent. Panel (c) shows the
  same mobile clusters (all colored red) and immobile clusters (all
  colored blue) to facilitate comparing their relative spatial
  distribution.}
\label{fig:cluster-pics}
\end{figure*}

We next provide a more complete account of the geometrical properties of
these clusters.  {Figure~\ref{fig:cluster-pics} provides a
  visualization of typical mobile and immobile particle clusters.}  To
quantify the structure of these clusters, we first consider the size
distribution $P(n)$ of the clusters at the characteristic times $t_M$
and $t_I$ of the mobile and immobile clusters, respectively.  This
distribution for mobile particle clusters has been previously examined
for a variety of systems, where it is appreciated that $P(n)$ can be
described by a power law with an exponential cut-off, namely
\begin{equation}
P(n) \sim n^{\tau_F}  \exp(n/n_0),
\label{eq:P(n)}
\end{equation}
where $n_0$ is proportional to $\langle n \rangle$.  {This
  distribution arises in the description of equilibrium branched
  polymers and clusters approaching a percolation transition (commonly
  referred to as ``lattice animals'').  We shall return to this analogy
  to help understand our findings.}  The Fisher exponent $\tau_F$ (using
standard notation from percolation theory~\cite{stauffer}) should not be
confused with a time scale.

\begin{figure}[b]
\includegraphics[clip,width=3.2in]{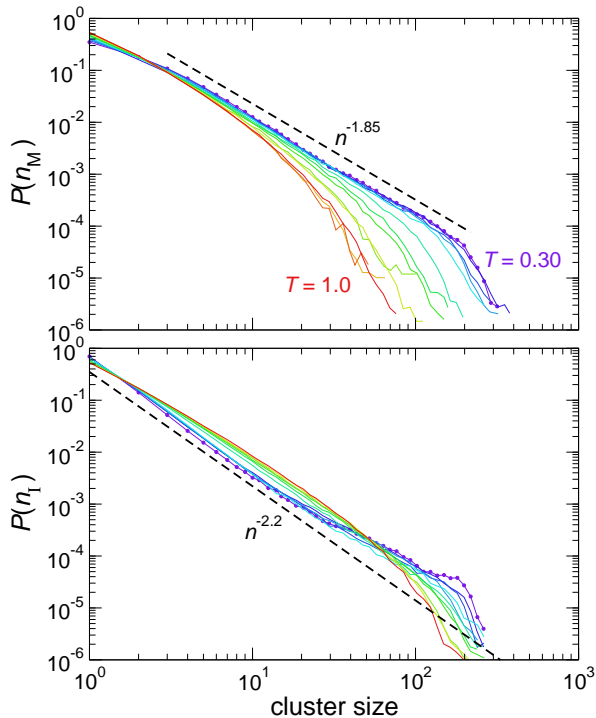}
\caption{The distribution of particle cluster sizes $P(n)$ for (a)
  mobile and (b) immobile particle clusters.  The distribution can be
  described by a power-law with an exponential cut-off, like clusters
  nearing the percolation transition. The dashed lines indicate a
  characteristic power-law.  Different colors represent different $T$,
  as in previous figures.}
\label{fig:cluster-dist}
\end{figure}

Fig.~\ref{fig:cluster-dist} shows that both the mobile and immobile
particle clusters follow eq.~\ref{eq:P(n)}, albeit with different
exponents $\tau_F$.  For mobile-particle clusters, we find $\tau_F =
1.85 \pm 0.1$.  Note that earlier work~\cite{gssg01} for this same model
indicated $\tau_F \approx 1.6$, but that work was limited to much
smaller clusters, and as a consequence was dominated by the behavior at
small $n_M$.  Our $\tau_F$ estimate is consistent with that for mobile
particle clusters in the Kob-Andersen binary LJ liquid ($\tau \approx
1.86$)~\cite{donati-pre}, and slightly larger than that for the
Kob-Andersen lattice gas model ($\tau \approx 1.6$).  All these $\tau_F$
estimates are smaller than found in percolation theory near the
percolation transition in 3D ($\tau_F = 2.18$)~\cite{stauffer}.  These
variations suggest that $\tau_F$ may be material dependent.  The mass
distribution $P(n)$ of the least-mobile particle clusters exhibits
similar scaling features to $P(n)$ for the most-mobile clusters, but the
exponent $\tau_F$ differs.  In particular, $\tau_F \approx 2.2$ is close
to that expected for percolation, although assignment of a precise
numerical value to $\tau_F$ for the least-mobile clusters is difficult,
given the present data.

{ We can better understand the value of $\tau_F$ by considering the
  possibility that mobile and immobile particle clusters are analogous
  to equilibrium branched polymers, which are directly related to
  percolation clusters.  In three dimensions, it is known that $\tau_F$
  ranges from 1.5 to about 2.2 for lattice animals~\cite{gaunt82,
    janse97, janssen12} and percolation
  clusters~\cite{naeem98,stauffer}, respectively, so that $\tau_F$ for
  branched polymers can be expected to be somewhat variable. This
  exponent reflects the effect of strong excluded volume interactions
  between and within these different types of model branched polymers.
  In mean field theory, which aplies above 8 and 6 dimensions for both
  lattice animals and percolation clusters, respectively, $\tau_F$ is
  exactly equal its classical Flory-Stockmayer value of 5/2.  Basically,
  lattice animals are swollen branched polymers and percolation clusters
  are branched polymers with screened excluded volume interactions so
  that these structures are branched polymer analogs of self-avoiding
  and random walk (more precisely, $\theta$-polymers) polymers
  describing equilibrium linear polymer chains.  In short, our exponent
  estimates for $\tau_F$ are consistent with the expected exponent range
  for branched polymers.  }

\begin{figure}[tb]
\includegraphics[clip,width=3.2in]{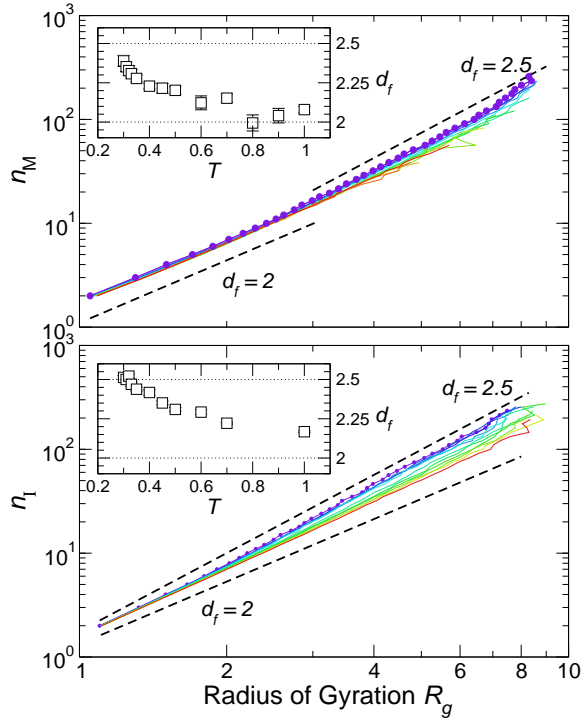}
\caption{Scaling of cluster mass $n$ with the radius of gyration $R_g$
  to define the fractal dimension $d_f$ of (a) mobile and (b) immobile
  particle clusters.  {The solid lines represent the
    simulation results, and different colors indicate different $T$, as
    in previous figures. Small circular symbols are shown for the lowest
    $T$, to indicate the density of data. The bold dashed lines are a
    guide to the eye, and provide approximate bounds on $d_f$.  This is
    more clearly seen in the insets, which show the $T$ dependence of
    $d_f$.  The dashed lines of the insets indicate the limiting
    behaviors discussed in the text.  }}
\label{fig:df-clusters}
\end{figure}

To further characterize the geometrical structure of these clusters at
their characteristic peak times, we examine the fractal dimension $d_f$
of the clusters defined by the scaling of cluster size
\begin{equation}
n \sim R_g^{d_f},
\end{equation}
where 
\begin{equation}
R_g^2 = \frac{1}{2N} \sum_{i,j} (r_i-r_j)^2
\end{equation}
is the radius of gyration, and $i$ and $j$ denote particles indices
within a given clusters.  Earlier work has suggested that $d_f \approx
2$ for mobile particle clusters~\cite{gbss05,lattice-animal-het}, which
corresponds to the value for lattice animals in 3D -- that is,
percolation clusters below the percolation threshold $p_c$. For
mobile-particle clusters, we indeed find that smaller clusters have $d_f
\approx 2$ (fig.~\ref{fig:df-clusters}).  However, for larger clusters,
which only occur for lower $T$, it appears the scaling crosses over to a
larger $d_f\approx 2.5$.  Since the appearance of these large clusters
occurs only at low $T$, extracting the best fit result for $d_f$ at each
$T$ results in the effective $d_f$ growing from roughly 2 to near 2.5 on
cooling (insets of fig.~\ref{fig:df-clusters}).  For immobile clusters,
the scaling of mass on $R_g$ for small and large clusters does not
noticeable change with size.  As in the case of mobile clusters, $d_f$
grows from 2 to near 2.5 on cooling.  Thus, there is significant
similarity in the geometrical structure of the mobile and immobile
particles clusters.  { However, the precise values of $d_f$
  should be taken with caution, since the range of the data covers less
  than a complete decade in $R_g$.}

{We can understand the changing value of $d_f$ by again considering
  the analogy to equilibrium branched polymers and lattice animals.
  Specifically, lattice animals are self-avoiding branched polymers with
  strong excluded volume interactions, and have a fractal dimension
  $d_f=2$ in three dimensions~\cite{stauffer, jack96}. The mobile and
  immobile particle clusters conform to this scaling at relatively high
  $T$, where they are sparse and not strongly interacting with each
  other.  Like the mass distribution exponent $\tau_F$, the fractal
  dimension $d_f$ of branched polymers is also sensitive to excluded
  volume interactions.  Branched polymers with screened, excluded volume
  interactions behave like percolation clusters, and have a fractal
  dimension $d_f \approx 2.5$ in 3D~\cite{stauffer}.  Thus, if we
  consider our clusters to be analogues of branched polymers, the
  increase of $d_f$ upon cooling can be interpreted as the result of
  screening of the interactions.  This situation is natural since, as
  the clusters grow upon cooling, the concentration describing the onset
  of their mutual interaction will decrease. We note that this crossover
  in exponent values has been anticipated for the branched polymer
  structures associated with Coniglio-Klein clusters in the Ising
  model~\cite{ws90}.

  The crossover behavior we observe for the exponent $d_f$ with
  temperature is not apparent in the size-scaling exponent $\tau_F$.  This
  might be understood from the fact that, in percolation, $\tau_F$ is far
  less sensitive than $d_f$ to changes in cluster structure or
  dimensionality~\cite{pzs01, stauffer}.  Given that the range of data
  used to determine $\tau_F$ and $d_f$ is rather modest, further study to
  determine whether these clusters can be exactly identified with
  branched equilibrium polymers is merited.  Additionally, examination
  of the anisotropy of clusters will be valuable to improve the
  comparison to branched polymers, since cluster shape is often a more
  discriminating metric of the cluster type than the size distribution or
  fractal dimension~\cite{family85}.}

\subsection{String-Like Cooperative Motion}

Mobile-particle clusters can be further decomposed into subsets of
string-like groups of cooperatively moving particles.  We now consider
the properties of these ``strings'', following an analysis parallel to
that just discussed for the clusters.  To identify string formed by
mobile particles, we follow the procedures originally developed in
ref.~\cite{strings-blj}. Specifically, using the same mobile particles
that we use to identify clusters, we consider two mobile monomers $i$
and $j$ to be in the same string if, over an interval $t$, one monomer
has replaced the other within a radius $\delta$.
Following ref.~\cite{strings-polymer}, which examined the same polymer
model, we choose $\delta =0.55$, although the results are not strongly
sensitive to this choice for reasonable values of $\delta$.  Since we
study a polymeric system, it is worth noting that the string-like
collective motion is not strongly correlated with chain
connectivity~\cite{strings-polymer}, so it should not be confused with
reptation-like motion.

\begin{figure}[tb]
\includegraphics[clip,width=3.2in]{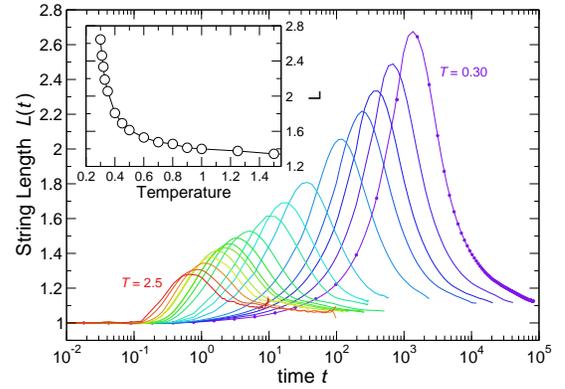}
\caption{The dynamical string length $L(t)$ for all $T$ studied.  The
  inset shows the $T$ dependence of the characteristic peak value,
  simply denoted as $L$.}
\label{fig:strings}
\end{figure}

For reference, we first show the average length (number of monomers) of
a string $L(t)$ for all $T$ studied (fig.~\ref{fig:strings}.  As
expected, $L(t)$ has a peak at a characteristic time which we label
$t_L$, and the time scale and amplitude of this peak grow on cooling,
indicating increased cooperative motion nearing $T_g$.  Since the
strings are subsets of the mobile particle clusters, the peak value of
$L$ is significantly smaller than that of the mobile particle clusters.
As a consequence, the percolation probability of the strings, even at
the lowest $T$ simulated, is negligible.

\begin{figure}[tb]
\includegraphics[clip,width=3.2in]{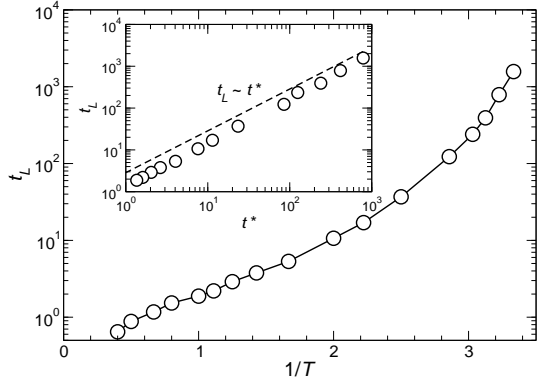}
\caption{The characteristic time $t_L$ of the peak string length.  The
  inset shows that, like the characteristic time of mobile particle
  clusters, $t_L$ scales nearly linearly with $t^*$.  
}
\label{fig:string-time}
\end{figure}

The characteristic time $t_L$ of the strings
(fig.~\ref{fig:string-time}) is similar to $t_M$ for the mobile particle
clusters, but is slightly larger, consistent with ref.~\cite{gvbmsg05}.
Moreover, like $t_M$, $t_L$ scales linearly with $t^*$
(fig.~\ref{fig:string-time} (b)).  Since $t^*$ scales linearly with the
characteristic diffusion time (appendix \ref{app:non-gaussian}), this
helps to clarify that the mobile particle time scales captured by the
clusters and strings relate to a diffusive relaxation time, rather than
the $\alpha$-relaxation time.  This time scale is naturally shorter than
$\tau_\alpha$ as a consequence of the breakdown of the Stokes-Einstein
relation.

\begin{figure}[b]
\includegraphics[clip,width=3.2in]{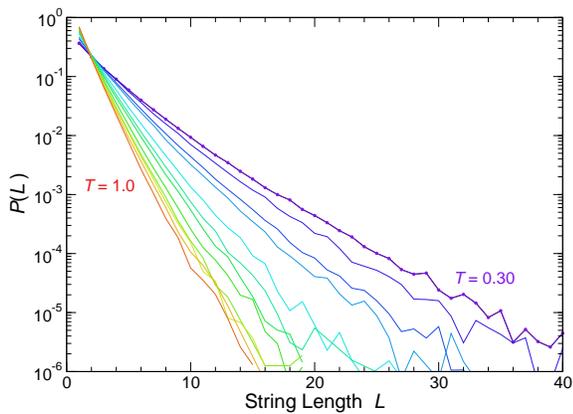}
\caption{The distribution of string lengths $P(L)$ follows an
  exponential law that can be anticipated by analogy with equilibrium
  polymerization~\cite{dfd06}.}
\label{fig:string-dist}
\end{figure}

\begin{figure}[tb]
\includegraphics[clip,width=3.2in]{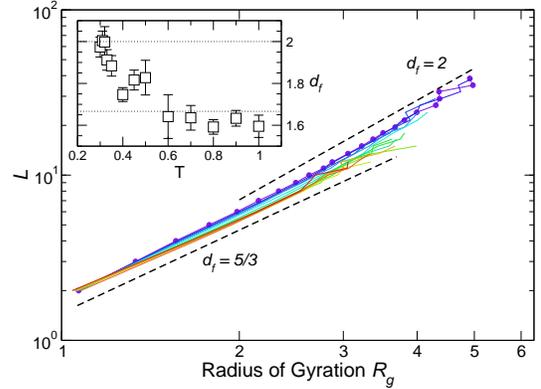}
\caption{The scaling $L \sim R_g^{d_f}$ to determine the fractal
  dimension $d_f$ of strings.  {The solid lines represent
    the simulation results, and different colors indicate different $T$,
    as in previous figures. Small circular symbols are shown for the
    lowest $T$, to indicate the density of data. The bold dashed lines
    are a guide to the eye, and provide approximate bounds on $d_f$.
    This is more clearly seen in the insets, which show the $T$
    dependence of $d_f$.  The dashed lines of the inset indicate
    expected limiting behaviors.  } Specifically, there appears to be a
  modest change in $d_f$ from $ \approx 5/3$ for short strings (the
  approximate value for a self-avoiding walk in 3D) to $\approx 2$ (a
  simple random walk, also characteristic of many branched polymers).
  Accordingly, a typical string at low $T$ are less extended, just as in
  polymer chains in dilute versus concentrated solutions.  }
\label{fig:df-strings}
\end{figure}

To complete the characterization of the strings, we examine the
distribution of string lengths $P(L)$ and their fractal dimension at the
characteristic time $t_L$.  As expected from earlier
works~\cite{strings-blj,strings-polymer}, fig.~\ref{fig:string-dist}
shows that $P(L)$ follows an exponential distribution that is
characteristic of linear equilibrium polymers~\cite{dfd06}.  To estimate
the fractal dimension $d_f$, we examine the scaling between $L$ and
$R_g$ in fig.~\ref{fig:df-strings}.  For short strings, we find that
$d_f \approx 5/3$, consistent with a self-avoiding walk in 3D.  For
longer strings, the scaling relation approaches $d_f=2$, the fractal
dimension of simple random walks or self-avoiding walks with screened,
excluded volume interactions~\cite{freed-book}.  {This screening
  effect has been seen in simulations of dynamically associating linear
  chain polymers~\cite{wittmer98}.} Since longer strings are prevalent
only at low $T$, the effective $d_f$ from fitting the entire range is
$T$-dependent, growing from $5/3$ to 2 on cooling (inset of
fig.~\ref{fig:df-strings}), reflecting an increased screening of
excluded volume interactions upon cooling as in the branched dynamic
clusters.  Hence, the strings appear to become somewhat more compact on
cooling.  It has been argued that cooperative motions should become
fully compact ({\it i.e.}\ $d_f\rightarrow 3$) at low $T$~\cite{ssw06},
but we see no indication of such a collapse for any of the cluster types
we have examined.  { As for our data for mobile and immobile
  particle clusters, the precise values of $d_f$ should be taken with
  caution, since the range of the data covers less than a complete
  decade in $R_g$.}

\begin{figure}[t]
\includegraphics[clip,width=3.2in]{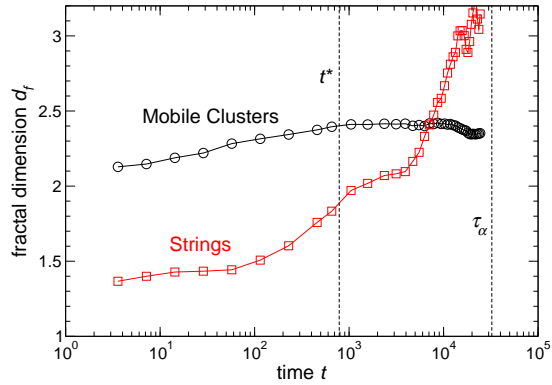}
\caption{The temporal evolution of fractal dimension $d_f$ of mobile
  clusters and strings at the lowest $T=0.30$. The mobile-particle
  cluster structure is relatively insensitive to $t$, but the string
  geometry differs dramatically between $t^*$ and $\tau_\alpha$.}
\label{fig:df-time}
\end{figure}

The time scale at which one examines cooperativity can be expected to be
important in the consideration of the cluster geometry.  Thus, we next
consider the fact that the geometrical structure of mobile clusters and
strings depends on the time scale on which one examines these objects.
Our previous analysis focused on $d_f$ at the characteristic peak time
of mobile particle clustering and string size, which is close to $t^*$,
a time that is significantly smaller (at low $T$) than $\tau_\alpha$.
Fig.~\ref{fig:df-time} shows the temporal evolution of $d_f$ for the
lowest $T$ studied.  While $d_f$ for mobile particle clusters is weakly
dependent on time, $d_f$ for the strings is indeed strongly dependent on
the time scale considered.  In fact, on time scales approaching the
structural relaxation time, the strings appear compact ($d_f \approx
3$), which may explain contradictory claims that cooperative motions
should form compact regions at low $T$~\cite{ssw06}.  On this long time
scale, the strings are quite small. This result emphasizes the fact that
it is critical to examine the cooperativity of motion on the appropriate
time scale, and thus quantification of these scales is necessary.

\section{Dynamical Scales and Relaxation}

\begin{figure}[t]
\centering
\includegraphics[width=3.2in]{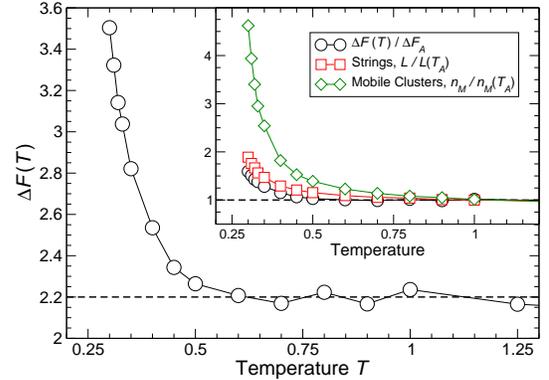}
\caption{Temperature dependence of the activation free energy $\Delta
  F(T)$, { evaluated from the relaxation time according to
    eq.~(\ref{eq:activation-energy}).}  The dashed line indicates the
  high $T$ asymptotic value $\Delta F_A = 2.2$.  The inset shows values
  of $\Delta F$, string mass $L$, and mobile cluster mass $n_M$
  normalized by their values at $T_A$ to facilitate comparison of the
  growth of these quantities on cooling. }
\label{fig:act-energy}
\end{figure}

A central challenge in describing glass formation is the origin of the
rapidly increasing relaxation time approaching $T_g$. This is the
defining characteristic of fragile glass-forming fluids.  If one makes a
natural assumption that relaxation is an activated process, transition
state theory~\cite{tst1941,tst1943,tst97} indicates a general Arrhenius
temperature dependence 
\begin{equation}
  \tau = \tau_0 \exp[\Delta F/T],
\label{eq:activation}
\end{equation} 
that is often observed in condensed phase relaxation process and in the
rates of chemical reactions.  At high $T$, $\Delta F(T) \equiv \Delta
F_A$, a constant, giving the widely known Arrhenius form.  At lower $T$,
this relation defines a generalized $T$-dependent activation free energy
\begin{equation}
\Delta F(T) = T \ln \, \tau/\tau_0,
\label{eq:activation-energy}
\end{equation} 
which we show for our data in fig.~\ref{fig:act-energy}.  This provides
a simple parametric description of the problem at hand: how can we
understand an activation barrier the grows on cooling to a value that is
several times larger than its high-$T$ limit?  Approaching $T_g$, this
growth typically reaches 4 to 8 times the high-$T$ limiting value
$\Delta F_A$, and the exponential nature of activation leads to
extremely large changes in relaxation.  The key element to explain the
increase of $\Delta F(T)$ is to recognize that such values cannot be
readily reconciled on the basis of single particle motion.  Both the AG
and RFOT approaches are built upon the notion that many particles are
involved in relaxation, and the scale grows on cooling toward $T_g$.
Accordingly, the change of $\Delta F(T)$ constrains any attempts to
explain the change in relaxation time of glass-forming liquids in terms
of a growing dynamical size scale.  Thus, as a simple starting point, we
consider the relative growth of $\Delta F$ with those of the cluster and
string sizes in the inset of fig.~\ref{fig:act-energy}.  We also make
the mathematically trivial, but conceptually important point, that the
existence of a fractional power law relation between $\tau$ and $t^*$
(fractional Stokes-Einstein relation) implies that the reduced
activation energy applies to {\it both} relaxation times, and indeed all
transport properties obeying such a power scaling relationship. This
explains why the AG model for diffusive relaxation can be equally well
applied to structural relaxation.

\subsection{Summary of the AG and RFOT Predictions}

The seminal work of Adam and Gibbs helped to establish a picture of
dynamics nearing $T_g$ where motion is dominated by
``cooperatively-rearranging regions'' (CRR), thereby introducing the
importance of a dynamical size scale.
Both the AG and RFOT approaches build on the activation picture for
dynamics.  For most fluids, the high $T$ ({\it i.e.} $T>T_A$) dependence
of relaxation is given by eq.~(\ref{eq:activation}), where $\Delta F
=\Delta F_A$.  AG associated this high $T$ activation barrier with
uncorrelated, single-particle motion.  On cooling toward $T_g$, AG argued
that motion becomes dominated by CRR, and that the barrier $\Delta F$ is
extensive in the number $z$ of rearranging monomers in a CRR, so that
\begin{equation}
\tau \sim \exp [z \Delta F_A/T].
\label{eq:AG-CRR}
\end{equation}
AG further argue that the fluid can be decomposed into $N/z$ such CRR, each
of which has a configurational entropy $s_{\rm conf}$, so that the total
\begin{equation}
S_{\rm conf} = \frac{N}{z}s_{\rm conf}.
\label{eq:AG-CRR-Sconf}
\end{equation}
Consequently, the relaxation can be directly relating $S_{\rm conf}$ via 
\begin{equation}
\tau \sim \exp [A/(TS_{\rm conf})],
\label{eq:AG-tau-Sconf}
\end{equation}
where the free energy $A$ subsumes previous constants.  This
configurational entropy picture has proved highly successful in
capturing the $T$ dependence of many supercooled fluids~\cite{gt67,
  ra98, rclc04, skt99,sastry01, speedy99, speedy01, hardspheres-ag,
  sslsss01, svsp04, mlsdst02}.  Unfortunately the CRR and $S_{\rm conf}$
are not explicitly defined by AG. Fortunately, numerous works have shown
that a potential energy landscape-based definition of $S_{\rm conf}$
appears robust~\cite{skt99,sastry01,speedy99, speedy01, hardspheres-ag,
  sslsss01, svsp04, mlsdst02}; other studies have shown that $z$ might
be defined in terms of string or mobile cluster
size~\cite{gbss03,gvbmsg05,sd11,pds13} -- a point we will examine in the
context of both AG and RFOT.

The similar RFOT description is built around a scaling description of
the problem~\cite{rfot89,lw07,bb04}.  RFOT theory proposes a ``mosaic''
picture, in which the liquid is divided into metastable regions with a
characteristic size $\xi$ (the ``mosaic'' length) -- conceptually like
the CRR idea of AG.  RFOT assumes the free barrier for
reorganization has a general scaling with size, so that
\begin{equation}
\Delta F \sim \xi^\psi.
\end{equation}
Accordingly, the implication is that $\tau$ scales as
\begin{equation}
\tau \sim  \exp [\xi^\psi/T]
\label{eq:RFOT-tau-xi}
\end{equation}
The free energy of this ``droplet'' is a balance between the entropic
contribution from the degeneracy of states $T S_c \xi^d$ (for a compact
droplet in dimension  $d$) 
and surface free energy, which should scale as $\Upsilon(T) \xi^\theta$,
where $\Upsilon$ is a generalized surface tension of the entropic
droplet, and the suface scaling exponent $\theta \le d-1$. If the
entropic droplet is not compact, such as is the case for our clusters
and strings, we can generalize this argument simply by replacing $d$
with $d_f$, as in the application of the droplet models to critical
phenomena.  The ordinary surface tension of many fluids is often found
to grow approximately linearly on cooling, and Cammarota {\it et
  al.}~\cite{ccggv09} argue that $\Upsilon$ of RFOT should grow at least
linearly on cooling.  Thus, assuming $\Upsilon(T) \sim T$ and balancing
the surface and volume effects yields a scaling between $\xi$ and
configurational entropy,
\begin{equation}
\xi \sim 1/S_{\rm conf}^{1/(d_f-\theta)}.
\label{eq:RFOT-entropy}
\end{equation}
Combining eqs.~\ref{eq:RFOT-tau-xi} and \ref{eq:RFOT-entropy} yields the
generalized AG-like relation between relaxation and entropy,
\begin{equation}
\tau \sim  \exp [A/(TS_{\rm conf})^{\psi/(d_f-\theta)}].
\label{eq:RFOT-tau-Sconf}
\end{equation}
{ We note that the concept of surface tension here is
  formally unclear, since there are no explicitly co-existing phases.
  However, subtle differences in the local packing of highly mobile and
  immobile particles certainly contributes an energy gradient near the
  interface of these regions.  Similarly, refs.~\cite{dsr10,zhang13}
  found that mobile-particle clusters found in a melting crystal can be
  identified with the nucleation of a fluid phase, so that the notion of
  a surface tension proper for mobile regions has a well defined meaning
  in this context. Further study in glass-forming liquids may illuminate
  the notion of the mobile particle clusters having a surface tension.
  Additionally, a more general scaling of $\Upsilon$ than simple
  proportionality to $T$ would lead to a slightly different scaling
  relation between $\xi$ and $S_{\rm conf}$.}

The values of the exponents $\psi$ and $\theta$ are not fixed in the
theory, {but on general physical grounds the exponents should obey
  the inequality $\theta \le \psi \le d-1$~\cite{fh88-2,bb04}.}  The
well-established entropy form of AG is recovered from RFOT provided that
$\psi = (d_f-\theta)$, leaving only one free exponent.  The original
presentation of RFOT by Kirkpatrick, Thirumalai, and
Wolynes~\cite{rfot89} presumes $d_f=d$ (compact droplets), and argues
that $\theta = d/2$ and $\psi = \theta$, satisfying the AG entropy form
{and exponent inequality}.  More recently, the exponents have been
examined in a number of computational and experimental
analyses~\cite{ccggv09, crz08, kds09}; these studies are inconclusive
regarding a universal value, but generally report values $\psi \approx
0.7$ to 1 and $\theta \approx 2$ to 2.3.  {These values are
  troubling, since, based on the exponent inequality, we expect the
  surface scaling exponent $\theta \le 2$ (for $d=3$), and further that
  $\theta$ should be larger than $\psi$.}

Both the AG and RFOT methods offer a way to relate the size or length
scale of motion with relaxation times, but do not directly specify how
this size scale should be measured.  Hence, we now consider if the
heterogeneity scales defined by mobile clusters or strings might be
appropriate for use within these theoretical descriptions.  However, we
note that, while many recent studies indeed focus on the size scales of
heterogenous motion in this context, there are reasons to be cautious,
and to consider other possible length scales~\cite{kds09, skds12}.

\subsection{Testing the Adam and Gibbs Approach}

\begin{figure}[tb]
\includegraphics[clip,width=3.2in]{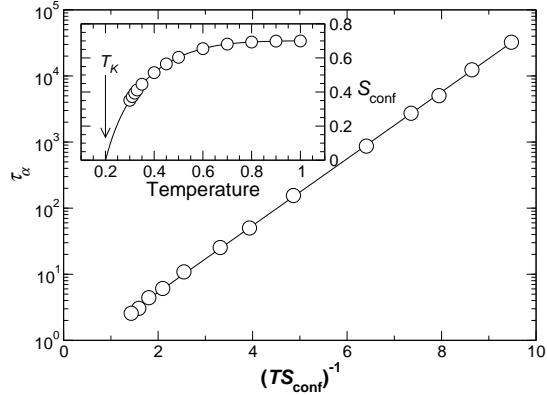}
\caption{Confirmation that the entropy representation of the AG theory
  (eq.~\ref{eq:AG-tau-Sconf}) is valid for the present system.  The inset
  shows the $T$ dependence of $S_{\rm conf}$ and the extrapolating
  vanishing temperature $T_K$, which is consistent with $T_0$ of the VFT
  fit (eq~\ref{eq:vft}) to $\tau$.}
\label{fig:AG}
\end{figure}

Although the relation between configurational entropy and relaxation
(eq.~\ref{eq:AG-tau-Sconf}) proposed by AG is not their starting point,
this is the most commonly tested and most broadly supported prediction
of the theory.  Hence we first wish to test whether this relation is
also valid in our system.  Evaluation of $S_{\rm conf}$ is a rather
cumbersome process, and so we describe the process fully in
appendix~\ref{appendix:sconf}. Fig.~\ref{fig:AG} verifies the validity
of eq.~\ref{eq:AG-tau-Sconf}.
The inset of Fig.~\ref{fig:AG} shows that extrapolating a simple
polynomial fit of $S_{\rm conf}$ to low $T$ yields ``Kauzmann''
temperature $T_K=0.20$ where $S_{\rm conf} \rightarrow 0$.  This is
exactly the same value $T_0$ obtained from independently fitting $\tau$
by the VFT function.  Hence, the vanishing of $S_{\rm conf}$ coincides
with the independently extrapolated divergence of relaxation time, a
comforting consistency check. 
{We note that $S_{\rm conf}$ is a relatively small contribution to
  the overall fluid entropy in comparison to the vibrational entropy of
  our polymer glass-forming liquid. This fact makes the experimental
  estimation of the difference between the total and vibrational
  contributions to the entropy particularly uncertain in polymer fluids,
  since there is no reliable means of estimating the vibrational entropy
  to high accuracy.}

We now continue to examine the proposal laid out by AG by considering
the relation of $\tau$ to a heterogeneity scale.  As described above,
the foundation of AG is that the activation energy for relaxation is
extensive in the mass $z$ of the CRR (eq.~\ref{eq:AG-CRR}).  Since the
CRR are not defined by AG, previous works have considered whether the
string mass $L$~\cite{gvbmsg05,sd11} or the mobile particle cluster mass
$n_M$~\cite{gssg01,gvbmsg05} might be appropriate measures.  For water,
it was shown that $n_M$ has the desired behavior, but over a relatively
limited range of $\tau$~\cite{gbss03}; for a simple spherically
symmetric model, both $L$ and $n_{\rm mobile}$ show the desired relation
to $z$, but over an even more limited range~\cite{gvbmsg05}.  More
recently, motivated by description of AG that CRR are the most basic
units of reorganization, refs.~\cite{sd11,pds13} found $L$ is an
appropriate measure of $z$, but did not consider cluster size $n_M$.
Unfortunately, none of these works could definitively exclude other
measures for $z$. It has also been appreciated that other length scales
associated with heterogeneity, such as from a four-point density correlation
function, would be too large at $T_g$ to be consistent with
$z$~\cite{kds10}.  We shall return to this point in the conclusion.

\begin{figure}[tb]
\includegraphics[clip,width=3.2in]{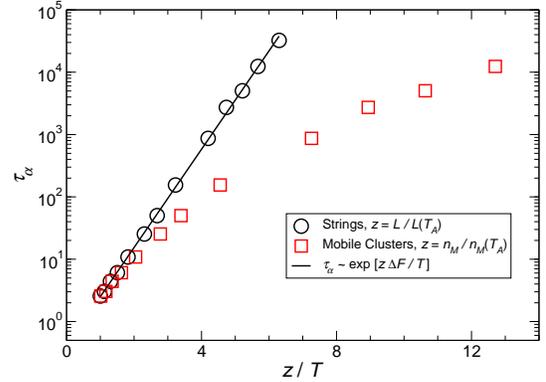}
\caption{Testing the mobile particles cluster and string sizes for
  consistency with the AG prediction that the activation energy is
  linear in the size $z$ of CRR.  The cluster size $n_M$ grows too
  rapidly on cooling to fit with the AG picture, the string size $L$
  appears consistent. See also figure~\ref{fig:act-energy} for a direct
  comparison of $L$ and $n_M$ to the relative change in $\Delta F$ with
  no free parameters in the comparison. }
\label{fig:AG-clusters}
\end{figure}

Here, we check the plausibility of both cluster and string size as a
measure of $z$ over a substantially broader range of $\tau$ than
previous works to provide improved clarity.  We thus consider
substituting for $z$ the peak string size $L/L(T_A)$ or the peak mobile
particle cluster size $n_M/n_M(T_A)$, where we normalize by the value at
$T_A$ so that $z$ has the expected value near unity for $T \ge
T_A$. Figure.~\ref{fig:AG-clusters} shows that, for $T\lesssim T_A$
$\log \tau$ is linear if we use $L$ as a proxy for $z$, but not using
$n_M$.  The deviation from cluster size is a consequence of the fact
that the mobile particle cluster size grows noticeably more rapidly than
the effective activation free energy, a fact already appreciated in the
inset of fig.~\ref{fig:act-energy}.  If AG is assumed correct at high
$T$, such an exponential relation should continue into the $T>T_A$
range, where $z$ should be near unity, or at least reach its asymptotic
value.  In such a case, the value of $\Delta F$ from the low $T$ fit
using eq.~\ref{eq:AG-CRR} should be comparable to the high $T$ estimate
of $\Delta F_A$ from an Arrhenius fit.  Using $z=L/L(T_A)$, we estimate
$\Delta F=1.8$, somewhat smaller that the value estimated from the
Arrhenius fit, $\Delta F_A=2.2$.  One interpretation of this discrepancy
is that using $L(T_A)$ as the normalizing factor is not entirely
correct, and a somewhat smaller value would be more appropriate.

\begin{figure}[tb]
\includegraphics[clip,width=3.2in]{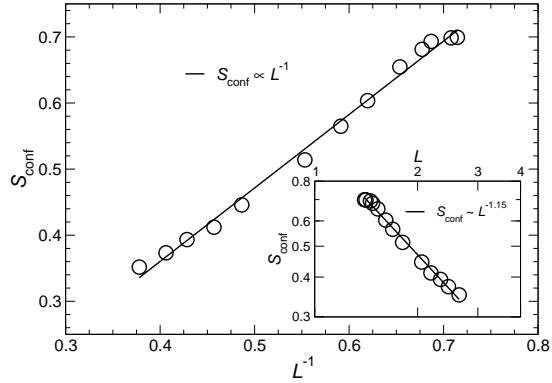}
\caption{Testing $S_{\rm conf} \propto 1/L$.  The inset checks for the
  best power-law relation, which yields an exponent slightly larger than
  one.}
\label{fig:CRR-Sconf}
\end{figure}

Given the apparent success of the string mass $L$ to describe $\tau$ and
the broadly reported validity of the entropy formulation of AG, we test
for consistency between these representations by checking the expected
relation $S_{\rm conf} \propto 1/z$ (fig.~\ref{fig:CRR-Sconf}). The data
support the linearity of the relationship, although there are some
systematic deviations at both the lowest and highest $T$.  This suggests
that, while $L$ captures the generally expected behavior of the CRR, a
more detailed refinement of the determination of $L$ may provide a more
accurate description of CRR.

The string length $L$ appears to be the most quantitatively valid choice
for the CRR, and is also consistent with the qualitative philosophy of
AG.  Specifically, recall that AG envision the CRR are dominated by the
{\it smallest} group of cooperatively moving monomers.  The physical
motivation for this is that probability of such a group will diminish
exponentially with the size, so that the smallest possible group that
allows for rearrangement will dominate the relaxation.  The strings are
both the smallest such unit, and also the only candidate in which all
particles move in a cooperative fashion.  While the particles of mobile
particles clusters are obviously spatially correlated, there is no {\it
  a priori} cooperativity in their displacements.  The strings are
precisely the manifestation of mobile particle cooperativity.  However,
we should be careful to point out that, while the qualitative language
of AG is appealing, in the end the quantitative predictability is the
most important measure.  We shall next explore which measure best
quantitatively fits with the formulation of the RFOT theory.

\subsection{Testing the RFOT Approach}

As discussed in the previous subsection, the validity of the entropy
formulation of AG dictates that $\psi = (d_f-\theta)$.  Hence there is
only one free exponent in the RFOT formulation.  We shall consider two
approaches to determine these exponents, which provides an internal
consistency check.

\begin{figure}[tb]
\includegraphics[clip,width=3.2in]{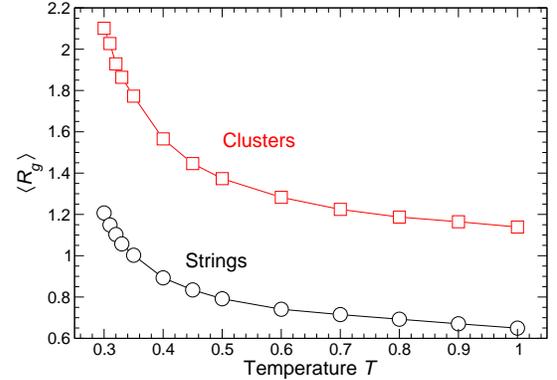}
\caption{The mean radius of gyration $\langle R_g \rangle$ for the
  strings and mobile particle clusters.  This defines a characteristic
  length that we can test within the RFOT framework.  }
\label{fig:Rg-temperature}
\end{figure}

A simple, but significant, difference between AG and RFOT is that RFOT
refers to a length scale of cooperative motion, rather than an extent or
mass of collective motion.  Consequently, to test whether any of the
clusters or strings might be appropriate, we need to consider a length
scale that defines the size of mobile clusters or strings.  The natural
length scale for these objects is the radius of gyration $R_g$ at their
respective characteristic times, which we show in
fig.~\ref{fig:Rg-temperature}.  Hence we can directly evaluate the
exponent $\psi$ from the scaling of $\tau$ with $\langle R_g \rangle$
(eq.~\ref{eq:RFOT-tau-xi}) for the strings and clusters.

\begin{figure}[t]
\includegraphics[clip,width=3.2in]{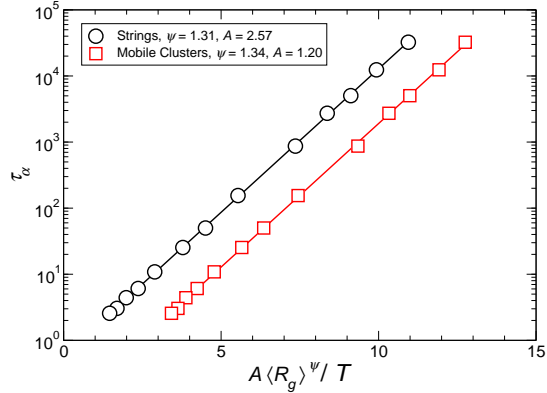}
\caption{Testing eq.~\ref{eq:RFOT-tau-xi} to determine the exponent
  $\psi$ of RFOT.  Both the mobile particle clusters and strings yield a
  consistent fit with $\psi \approx 1.3$.  The data for the mobile
  particle clusters are shifted by two units along the abscissa for clarity of the figure.  
 }
\label{fig:RFOT-psi}
\end{figure}

Figure~\ref{fig:RFOT-psi} shows the scaling of $\tau$ with $\langle R_g
\rangle$ for the strings and clusters, from which we obtain with the
best fit for the exponent $\psi$. Given our previous findings for the AG
approach indicating that string mass relates to $\tau$ while the cluster
mass does not, we would expect a superior fit for the strings.  Instead,
we find that the exponent $\psi \approx 1.3$ for strings and clusters is
identical within the limits of our determination.  Essentially, this is
a consequence of the fact that the $T$-dependence of $R_g$ is nearly the
same for strings and for clusters, while the $T$-dependence of cluster
mass differs noticeably.  This apparent paradox can be resolved by
recognizing that the largest dimension of a cluster dominates $R_g$, so
that systems with different mass can have similar $R_g$.  This is
consistent with the expectation that the largest dimension of mobile
clusters is associated with long, string-like cooperative regions.

\begin{figure}[t]
\includegraphics[clip,width=3.2in]{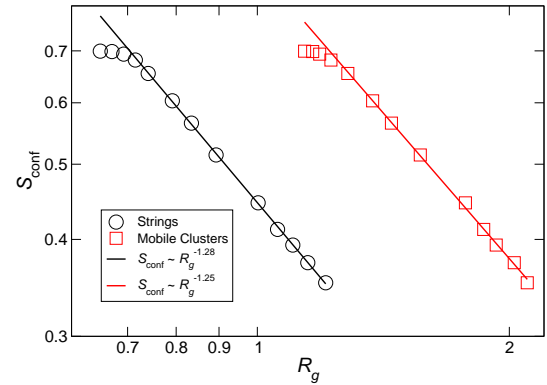}
\caption{Evaluation of the surface scaling exponent $\theta$ from the
  scaling law eq.~(\ref{eq:RFOT-entropy}).  The value $d_f-\theta \approx
  1.3$ is consistent with $\psi$, and with the expectations from AG. }
\label{fig:Rg-Sconf}
\end{figure}

We next check for consistency of the value $\psi$ with the expectation
that $\psi = d_f-\theta$.  We can independently determine $d_f-\theta$
from the scaling of $S_{\rm conf}$ with $R_g$
(eq.~\ref{eq:RFOT-entropy}), as shown in fig.~\ref{fig:Rg-Sconf}.  While
the data deviate from a power-law at high $T$, the lower $T$ data
indicate $d_f-\theta \approx 1.3$, consistent with our estimates of
$\psi$.  The success of these independent approaches significantly
increases our confidence in these estimates.  Based on our previous
findings for $d_f$ for strings, we can also estimate $\theta = 0.3$ to
$0.7$.  { The value of $\theta$ is small in comparison with
  values estimated refs.~\cite{ccggv09, crz08, kds09}.  However, a small
  value for $\theta$ is physically plausible.  For example, in the Ising
  spin glass, a direct evaluation yields $\theta \approx 0.2$ to 0.35 in
  three dimensions~\cite{fh88-2,mcmillan84,bm84}.  Moreover, our value
  obeys the expected inequality $\theta \le \psi \le d-1$, which the
  previous estimates violate~\cite{ccggv09, crz08, kds09}.}

While both string and mobile particle cluster sizes demonstrate
reasonably scaling within RFOT, it appears the success of the cluster
description is dependent on the limiting dimension dictated by the
string size.  The strings also appeared to be the only reasonable
description of CRR within the AG framework.  Hence, we can find a
satisfying consistency for both the AG and RFOT descriptions using the
strings as a measure of CRR or mosaic scale, where the exponent values
of RFOT are constrained to satisfy the formulation of AG.  

\section{Discussion and Conclusion}

We have examined the geometrical structure of clusters and string-like
cooperative motions in a model glass-forming polymer melt.  We found an
aesthetically pleasing symmetry in the geometry of high- and
low-mobility clusters, {\it i.\ e.}, they both conform to statistical
geometry of equilibrium branched polymers.  In doing so, we also
developed a novel method to identify low-mobility particles based on
persistent caging.  Most importantly, we have examined the question of
whether these heterogeneity scales can be identified with the scales
anticipated by the AG and RFOT descriptions of glass formation.  We
found the strings apparently provide the most consistent description
of the CRR or mosaic length described by these respective theories.

An important observation arising from our work is that these different
quantifications of heterogeneous dynamics in fact correspond to distinct
relaxation time scales, and therefore distinct processes of importance
in the relaxation of a glass former.  In other words, there is no single
or unique heterogeneous dynamical scale in the system. These immobile
particles are apparently related with the breakdown of the linear
scaling between diffusive and viscous relaxation, while the fragility of
glass-formation is apparently more related to string-like cooperative
motion~\cite{sd11}. Thus, conventional wisdom regarding the role of
heterogeneous dynamics on typical aspects of glassy behavior, requires
further examination.  Apparently, there is no single dynamic
heterogeneity scale in glass-forming liquids.

There is further evidence of these scales and their significance on
condensed matter relaxation. For example, recent simulations of
superheated Ni crystals~\cite{zhang13} also find a large increase of the
non-Gaussian parameter, mobile particle clusters, and string-like
collective motion.  However, in this system, there are no immobile
particle clusters of finite extent present; there are only immobile
atoms in a crystal lattice, and mobile particles having the usual
constituent strings.  Significantly, there is no decoupling of
structural relaxation from the self-intermediate scattering function and
the diffusion coefficient, nor any stretched exponential decay of the
self-intermediate scattering function in the superheated crystal.  Similarly,
in a recent study~\cite{skds12,shiladitya-seviolation} on the dependence
of dynamics on spatial dimensionality, it is found that the degree of
Stokes-Einstein breakdown decreases while the fragility paradoxically
increases with spatial dimensionality for dimensionality greater than
two.  The findings of these studies suggest that the immobile particle
clusters, rather than mobile particles, are primarily responsible for
decoupling and stretched exponential stress relaxation of glass-forming
liquids.  This possibility merits systematic study and points to the
different types of heterogeneity (mobile and immobile particles) having
significantly different impacts on the fluid dynamics. In other words,
dynamic heterogeneity comes in different types that must be properly
discriminated.  

As we alluded to earlier, another common approach to extract a length
scale for heterogeneity is via a four-point correlation function. Proper
determination of the four-point scale $\xi_4$ can be strongly affected
by finite size~\cite{kds10}, but careful extractions have show that
$\xi_4$ grows more rapidly than would be expected for the CRR of the AG
theory~\cite{kds09}.  It is possible that $\xi_4$ could be consistent
with the mosaic scale of the RFOT theory, but we expect a single measure
should be compatible with both approaches, since they are largely
complementary. The reason for the difference in the scaling of $\xi_4$
with that observed for the strings can be readily understood by
considering their characteristic times; $\xi_4$ is determined at the
time of the peak in the four-point susceptibility, which has essentially
the same temperature dependence as $\tau_\alpha$.  As we have shown,
$\tau_\alpha$ is also the time scale of immobile particles, and is
distinct from the time scale (and hence length scale) of mobile
particles and strings.  Accordingly, $\xi_4$ is primarily sensitive to
particles of low mobility~\cite{weitz06}.  This is a consequence of the
fact that, in defining the four-point function, a particle size $a$ is
introduced to limit the effects of vibrational motion, and is chosen to
be larger than the typical cage size, following ref.~\cite{lssg03}. The
choice of $a$ controls the scale of relaxation associated with the
four-point function.  Choosing a smaller value of $a$, closer to the
cage size, for example, should lead to a measure of heterogeneity on a
smaller time scale, perhaps similar to $t^*$; such a choice would
presumably be more sensitive to string-like excitations.  Efforts in
this direction, along with other approaches to extract the size scale of
string-like cooperativity, would be valuable to better understand the
findings of the present paper within a more traditional liquid-state
correlation function approach.

In conclusion, the analysis of the relationship between the various
clusters in the context of the AG and RFOT theories reveals that the
strings are a particularly good candidate for the CRR of AG
theory. While a similar conclusion may also be reached in the context of
RFOT, the fact that all the heterogeneous clusters considered here show
fractal structure with a fractal dimension lower than $d = 3$ makes a
conclusive comparison difficult at this time, since the mosaic picture
in the RFOT framework normally assumes the rearranging regions to be
compact.  Consequently, it will be valuable to revisit the formulation
of RFOT, as the suggested fractal nature of the entropic droplets has
implications for the concept of the effective surface tension of these
regions.

\appendix
\def\thesection{Appendix \Alph{section}}

{
\section{Non-Gaussian and Diffusive Time Scales}
\label{app:non-gaussian}
\def\thesection{\Alph{section}}
The non-Gaussian parameter $\alpha_2(t)$ is often used to quantify the
deviation of particle or segmental displacements from the Gaussian
distribution expected for simple fluids.  The maximum deviation occurs
at a characteristic time $t^*$, and it is well known that $t^*$ is
smaller than $\tau_\alpha$, and has a weaker temperature dependence than
$\tau_\alpha$.  It is then natural to wonder to what physical process
$t^*$ relates.  

Combining the Stokes-Einstein relation for spheres
\begin{equation}
D=\frac{k_BT}{6\pi \eta R_h},
\end{equation}
where $D$ is the diffusion coefficient, $\eta$ is the fluid viscosity,
and $R_h$ the particle hydrodynamic radius, with Maxwell's relation
$\tau = \eta/G_\infty$ (where $G_\infty$ is the high-frequency shear
modulus) leads to
\begin{equation}
\frac{D}{T} \propto \frac{1}{\tau}.
\end{equation}
In other words, the reduced diffusion coefficient $D/T$ should define an
inverse relaxation time. Given the relative slow variation of $t^*$ with
$T$ compared to $\tau_\alpha$ from the intermediate scattering function,
we check whether $t^*$ can be identified with a diffusive relaxation
time defined in this way.  Due to the polymeric nature of our system,
$D$ cannot be readily evaluated, since mean-squared displacement will
only be linear on a much longer time scale, associated with the chain
center-of-mass diffusion; this requires chain displacements at least on
the order of the chain radius of gyration, more than we can readily
simulate at low $T$. However, we can check for a relation between $D/T$
and $t^*$ for the Kob-Anderson binary Lennard-Jones fluid~\cite{ka95-1},
the most commonly studied computational glass-forming system, where low
$T$ data is accessible.

\begin{figure}[tb]
\begin{center}
\includegraphics[clip,width=3.2in]{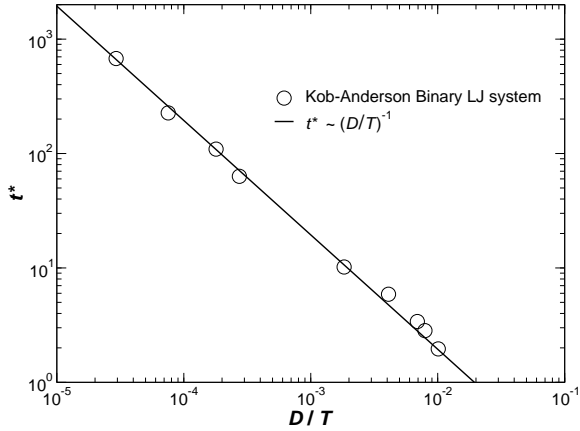}
\end{center}
\caption{Linear scaling between $t^*$ and the (inverse) characteristic
  diffusion time $D/T$ in the Kob-Anderson binary LJ
  mixture~\cite{ka95-1}, demonstrating that the time scale $t^*$
  corresponds to the time scale for mass diffusion. }
\label{fig:D-tstar}
\end{figure}

Figure~\ref{fig:D-tstar} shows that the (inverse) characteristic
diffusion time $D/T$ is linear with $t^*$ for the entire range of data,
covering several decades.  This result clarifies that $t^*$ can be
associated with a diffusive time scale.  Given the known ``decoupling''
of structural relaxation and $D/T$ from the Stokes-Einstein relation, we
accordingly expect the same decoupling between $t^*$ and $\tau_\alpha$
in our polymer system, as observed in fig.~\ref{fig:fqt+alpha2}.

It should be appreciated that the decoupling relation between $D$ for
the overall chain displacements of a polymer and the long-time
shear-stress relaxation time can exhibit a separate relationship from
$t^*$ and $\tau_\alpha$~\cite{sok-schweiz}.  This is a consequence of
the fact that heterogeneity at the scale of the chain radius of gyration
can differ from heterogeneity at the monomer or segmental scale. Thus,
in the polymer system, $t^*$ should be  thought of as relating to a
local monomer diffusive process, and $\tau_\alpha$ to a segmental
structural relaxation time.  }

\section{Immobile Particle Definition}
\label{appendix:immobile}

\begin{figure}[t]
\includegraphics[clip,width=3.2in]{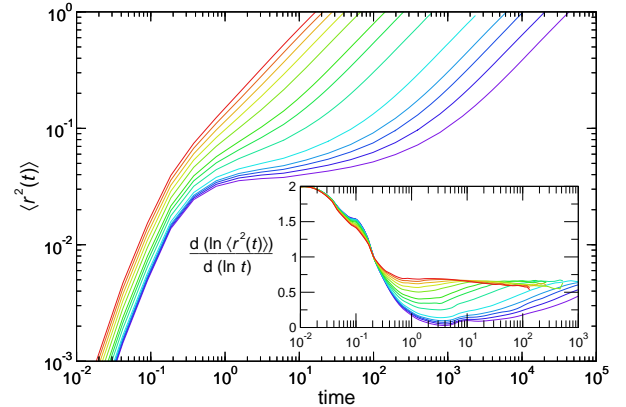}
\includegraphics[clip,width=3.2in]{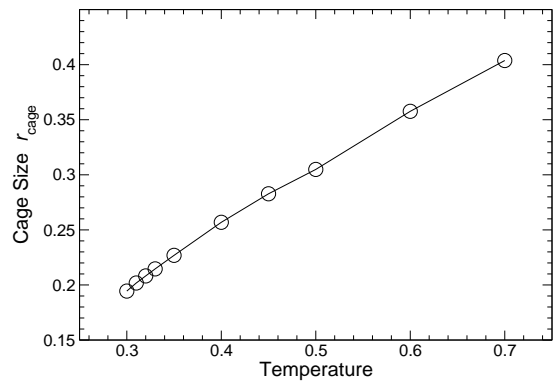}
\caption{(a) Identification of the cage size from the mean-squared
  displacement.  The inset shows the logarithmic derivative, which shows
  a minimum that we use to define $\langle r^2_{\rm cage}\rangle$.  (b)
  The resulting $T$ dependence of the cage size $\langle r^2_{\rm
    cage}\rangle^{1/2}$. Particles with a squared displacement less than
  $\langle r^2_{\rm cage}\rangle$ are defined as ``caged particles''.}
\label{fig:cagesize}
\end{figure}

To study the structure of highly immobile particles, we need to devise a
physically sensible algoritm that picks out an appropriate subset of
low-mobility particles.  Since the caging of particles by their
neighbors is one of the hallmarks of glass formation, we utilize the
concept of ``caged particles''.  To do so, we must identify the cage
size.  We can formally do this via the mean-squared displacement
$\langle r^2(t) \rangle$.  Fig.~\ref{fig:cagesize}(a) shows that
$\langle r^2(t) \rangle$ has a plateau at a characteristic size in the
approximate range 0.04 to 0.09 (for a cage radius 0.2 to 0.3).  To
precisely define the cage size, we take advantage of the fact that the
logarithmic derivative $d(\ln\langle r^2(t) \rangle) / d(\ln t)$
exhibits a clear minimum on the time scale of particle caging, $t_{\rm
  cage}$.  We thus define the cage size by $r_{\rm cage} \equiv \langle
r^2(t_{\rm cage}) \rangle^{1/2}$. We show the $T$ dependence of $r_{\rm
  cage}$ in fig.~\ref{fig:cagesize}(b) for $T\lesssim T_A$; at higher
$T$, $\langle r^2(t) \rangle$ transitions from ballistic motion
($d(\ln\langle r^2(t) \rangle) / d(\ln t)= 2$) to sub-diffusive motion
($d(\ln\langle r^2(t) \rangle) / d(\ln t) \approx 0.6$ without
intervening particle caging.  The sub-diffusive behavior is well known
for this model~\cite{basch-soft-rev10}, arising from polymeric effects.

\begin{figure}[t]
\includegraphics[clip,width=3.2in]{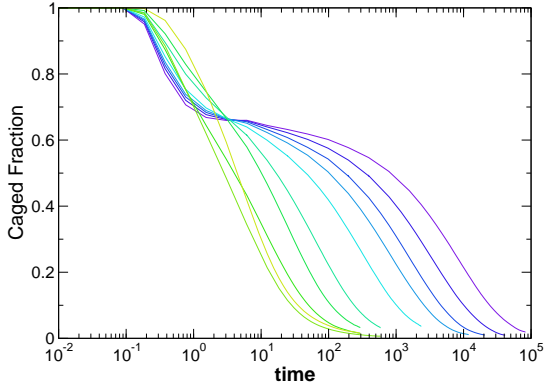}
\caption{Dynamical fraction of caged particles for $T<T_A$.}
\label{fig:caged-fraction}
\end{figure}

Having unambiguously defined a cage-size, we proceed to track the
behavior of caged particles ({\it i.~e.}\ particles with displacement
less than $r_{\rm cage}$). Fig.~\ref{fig:caged-fraction} shows
the fraction of caged particles, which, as expected, decreases with
time.  We note that, formally, this fraction is identical to the
``self-overlap'' $Q_s(t)$ used in the four-point correlation function
formalism~\cite{lssg03}, although the particle ``size'' used is
typically fixed at a value 0.3, independent of $T$, and substantially
larger than the cage size at low $T$.

Since we wish to understand the tendency for these caged particles to be
spatially correlated, we evaluate the average cluster size of these
immobile particles.  However, we must take into account the fact that
the number of these caged particles decreases with time, and thus there
is a trivial effect on the cluster size of the number of caged
particles.  To remove this trivial effect, we normalize the caged
particle cluster size by the cluster size of the same fraction of
particles chosen at random; this allows us to see how the tendency to
cluster compares to the random case, independent of the nature of the
underlying dynamics. 

\begin{figure}[b]
\includegraphics[clip,width=3.2in]{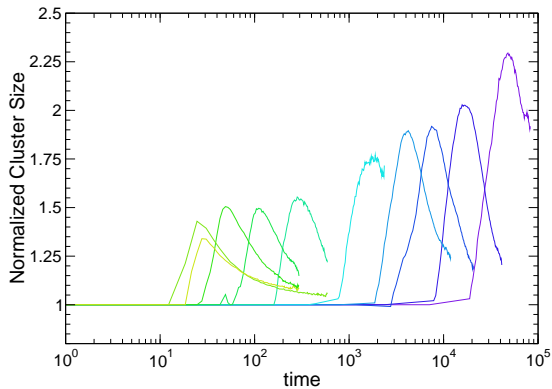}
\caption{The normalized cluster size for caged particles for $T<T_A$,
  where the cage is well-defined.}
\label{fig:caged-clusters}
\end{figure}

Figure~\ref{fig:caged-clusters} shows the normalized cluster size of the
caged particles as a function of time.  The qualitative behavior matches
what we observe for other dynamical clusters: as smaller times, the
effect of clustering is weak, and there is a characteristic time where
the cluster size reaches a peak.  As discussed in the main text, this
characteristic time is similar to the $\alpha$-relaxation time.  This
time, along with the peak size of the caged-clusters, define the
characteristic features that we wish to capture.

\begin{figure}[tb]
\includegraphics[clip,width=3.2in]{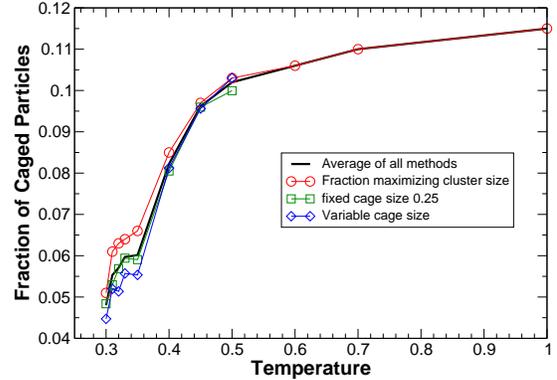}
\caption{The characteristic fraction of low mobility particles for
  several different approaches.  All approaches yield similar fractions
  for all $T$.}
\label{fig:immob-fraction}
\end{figure}

To simplify the analysis of immobile particles and draw a parallel to
the analysis of the mobile particles (where there is a fixed,
$t$-independent fraction of particles considered), we consider a
simplification of this approach that still captures the characteristic
peak time and amplitude of caged particles.  Specifically, if we look at
the characteristic peak time of the caged particles, we can identify the
fraction of caged particles at this time, which we show in
fig.~\ref{fig:immob-fraction}.  This characteristic fraction of caged
particles increases with $T$, as the mobility subsets become less
distinct at higher $T$.  For all subsequent analysis, we use this
$T$-dependent fraction for all time, to parallel the approach for the
mobile particles. By construction, this fixed fraction reproduces the
characteristic time and size the our time-dependent fraction of
caged-particles reveals, so that we do not alter these important
features.

A natural concern is the sensitivity of this approach to the definition
of the cage size, since this is the only parameter that must somehow be
chosen.  To test this, we also considered a $T$-independent cage size
$r_{\rm cage}$ of 0.2 or 0.25.  We find that these $T$ independent sizes
of course yield quantitatively different results, but that the time and size
scales of the immobile particles  all scale in the same way.  Additionally,
as shown in fig.~\ref{fig:immob-fraction}, the characteristic fraction
of caged particles is nearly the same as our definition based on a
$T$-dependent cage size, showing the precise definition of cage size
does not strongly affect the characteristic immobile fraction.

We also considered another approach to extract the low mobility subset.
This approach is motivated by ref.~\cite{gssg01}, where they determined
a characteristic fraction of highly mobile particles by finding the
fraction that maximizes the cluster size formed by those highly mobile
particles relative to same fraction of particles chosen at random.  This
measure provides a way to capture the fraction of particles that most
strongly exhibit spatial clustering, without consideration of the
underlying mobility distribution. We use the same approach, but finding
the fraction of least mobile particles that maximizes the relative
clustering.  We show this characteristic fraction together with our
estimates from the cage size in fig.~\ref{fig:immob-fraction}, and find
that we recover nearly the same fraction by this approach.  Apparently,
the characteristic fraction of immobile particles is not strongly
sensitive to the exact choice of parameters, providing confidence in the
robustness of our analysis.  For the calculations presented in the main
body of the manuscript, we simply use the mean of all these estimates.

Finally, we point out that in all approaches to identify low mobility
particles, there appears to be an unanticipated feature at $T\approx
0.35$, previously estimated as the characteristic $T_c$ for this system
at this density~\cite{ssdg02}.  This is a curious result, but at present
we have no explanation for such behavior.

\section{Calculation of Configurational Entropy}
\label{appendix:sconf}
\def\thesection{\Alph{section}}

Our goal is the evaluation of the configurational entropy $S_{\rm
  conf}$, which enumerates the density of stable potential energy minima
sampled by the melt at equilibrium.  Procedures for evaluating $S_{\rm
  conf}$ have been developed applied to a variety of systems, including
water~\cite{sslsss01}, binary LJ mixtures~\cite{skt99,sastry01},
silica~\cite{svsp04}, and orthoterphenyl~\cite{mlsdst02}.  We follow a
similar procedure, whereby the overall entropy can be partitioned into
vibrational $S_{\rm vib}$ and configurational components, i.e.\
\begin{equation}
S = S_{\rm conf} + S_{\rm vib}.
\end{equation}
Our approach will be to evaluate directly $S$ and $S_{\rm vib}$, and by
their difference $S_{\rm conf}$.

\subsection{Total Entropy}
To evaluate the absolute entropy of the polymer, we employ the
thermodynamic integration technique~\cite{frenkel-book}.  In this
method, the free energy is calculated by parametrically coupling the
potential energy of the system $U_{\rm poly}$ to the potential energy
$U_{\rm ref}$ of a reference system for which the free energy can be
directly, analytically evaluated.  The coupling potential is normally of
the form
\begin{equation}
        U(\lambda)=(1-\lambda)U_{\rm polymer}+\lambda U_{\rm ref}
\label{eq:therm-int1}
\end{equation}
where the coupling parameter $0\le \lambda \le 1$.  The free energy can
then be evaluated by 
\begin{equation}
        F_{\rm poly}=F_{\rm ref}-\int_{0}^{1}\left\langle
        U_{\rm ref}-U_{\rm poly}\right\rangle  d\lambda.
\label{eq:therm-int2}
\end{equation}

This procedure is complicated by the FENE potential that bonds nearest
neighbors, since it diverges as the bond length approaches $R_0$.  The
normally chosen reference potentials do not restrain the bond length,
and so the contribution from the FENE potential diverges strongly as
$\lambda \rightarrow 1$.  The avoid this complication, we perform a
``two-step'' thermodynamic integration, where we first perform an
integration from the FENE potential to a harmonic bond potential
\begin{equation}
U_{\rm harm} = \frac{k_{\rm harm}}{2} (r-R_1)^2
\end{equation}
that does not exhibit the strong divergence as bond length grows; when
then perform an integration from the harmonically bonded polymer to the
reference system.  We choose $k_{\rm harm} = 980$ and $R_1=0.87$ so that
the location of the minimum and the curvature at the minimum are very
near to that of the FENE bond (when combined with the core LJ
repulsion).  While the free energy for the harmonically bonded polymer
is not analytically known, it is not needed, as it drops out in the
final expression for the free energy
\begin{equation}
\begin{multlined}
        F_{\rm poly}=F_{\rm ref}
       -\int_{0}^{1}\left\langle U_{\rm polymer}-U_{\rm
           harm}\right\rangle  d\lambda_1 \\
       -\int_{0}^{1}\left\langle U_{\rm ref}-U_{\rm harm}\right\rangle  d\lambda_2.
\end{multlined}
\label{eq:therm-int3}
\end{equation}
        
For the reference potential, we use a potential that shares the system
periodicity~\cite{vs11}, namely
\begin{equation}\label{eq:Uref}
        U_{\rm ref}(r)=-U_{0}\sum_{i=1}^{3}\cos\left(\frac{2\pi}{L}r_{i}
 \right),
\end{equation}
where $U_{0}=10$ is the amplitude of the potential, $L$ is the length of
the container and $r_{i}$ is the coordinate of a particle in direction
$i$. For an $N$ particle system interacting through $U_{\rm ref}$,
evaluation of the partition function shows that
\begin{equation}\label{eq:Free_ref}
        F_{\rm ref}=\frac{3N}{\beta}\ln\left(\frac{\rho^{\frac{1}{3}}\Lambda}{\rm I_{0}(\beta U_{0})}\right).
\end{equation} 
Here $\rho$ is the number density and $\rm I_{0}\left(x\right)$ is the
modified Bessel function of the first kind. Combining with
Eq.~(\ref{eq:therm-int3}), we have $F$ for the our system at some fixed
temperature $T_{0}$ and density. Accordingly, we can evaluate the
entropy for a reference $T_0$,
\begin{equation}\label{eq:S_for_T0}
 S(T_{0})=\frac{ E(T_{0})-F(T_{0})}{T_{0}}.
\end{equation} 
We obtain $S$ for any $T$ by exploiting the fact that
\begin{equation}\label{eq:CV}
C_{V}=T\left(\frac{\partial S}{\partial T}\right) _{V}=\left( \frac{\partial E}{\partial T}\right) _{V},
\end{equation} 
so that 
 \begin{equation}\label{eq:S_for_any_T}
S(T)=\int_{T_{0}}^{T}\frac{1}{T}\left( \frac{\partial
E(T)}{\partial T}\right) _{V}dT+S(T_{0}) .
\end{equation} 
The integrand can be evaluated numerically from data for $E(T)$.  Since
we must explicitly include Planck's constant $\hbar$, we select units
appropriate for a monomer of a typical polymer, like polystyrene;
specifically, we choose $\epsilon = 1$~kJ/mol, $\sigma=1$~nm, and $m =
100$~g/mol; using these units, $\hbar = 0.0635078$~kJ$\cdot$ps/mol.

\begin{figure}[tb]
\includegraphics[clip,width=3.2in]{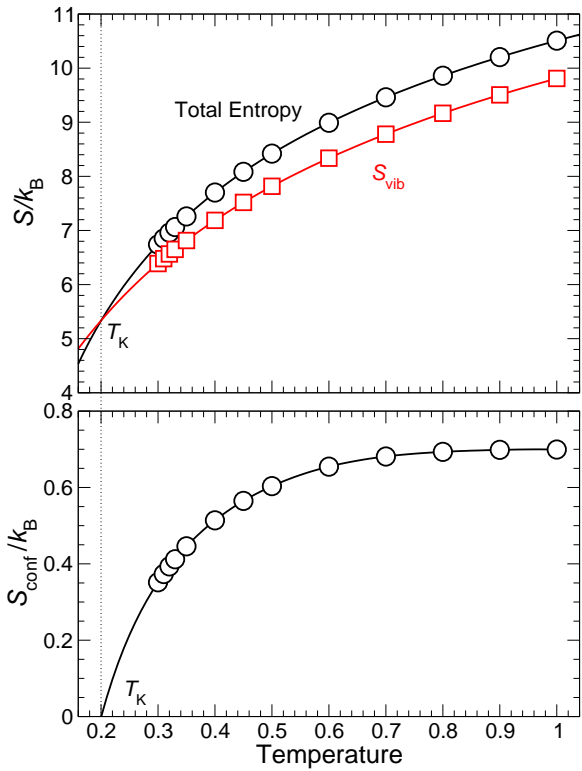}
\caption{(a) The total entropy $S$ and vibrational entropy as a function
  of temperature.  (b)  The configurational component $S_{\rm conf} =
  S-S_{\rm vib}$. The line is a guide for the eye. }
\label{fig:entropies}
\end{figure}

\subsection{Vibrational Entropy}

The vibrational component of the entropy $S_{\rm vib}$ reflects the
contributions of the basin shape to the vibrational behavior.  We can
partition 
\begin{equation} 
S_{\rm vib} = S_{\rm harm} + S_{\rm anh}
\end{equation}
In the harmonic approximation, the basin entropy
\begin{equation}
S_{\rm harm} = k_B \sum_{n=1}^{3N-3} \left[ 1 - \ln \left( \frac{\hbar
      \omega_n}{k_B T} \right) \right]
\end{equation}
where ${\omega_n}$ are the normal modes associated with the basin
minimum.  To evaluate ${\omega_n}$, we must first determine the basins
associated with the equilibrium liquid.  To do so, we perform a
conjugate gradient minimization of the potential energy, starting from
instantaneous snapshots of the equilibrium polymer; we locate the
corresponding minimum, or inherent structure (IS), within a numerical
tolerance of $10^{-15}$.  Using the configuration at the minimum, we
evaluate the Hessian matrix
\begin{equation}
H_{ij} = \frac {\partial V}{\partial \vec{r}_i \partial \vec{r}_j},
\end{equation}
the matrix of the curvatures of the potential energy.  In the harmonic
picture, the eigenvalues $\lambda_n = m \omega_n^2$, so we directly obtain
the normal modes $\{\omega_n\}$.  For each $T$, we generate the IS and
$\{\omega_n\}$ for at least 100 configurations that are well-separated
temporally.

The anharmonic contribution to $S_{\rm vib}$ for many systems is
negligible.  To check the anharmonic contribution $S_{\rm anh}$, we
first consider and anharmonic energy
\begin{equation}
E_{\rm anh}(T) = E-\frac{3}{2} Nk_BT-e_{\rm IS}
\end{equation} 
where $e_{\rm IS}$ is the inherent structure energy and $3/2\; Nk_BT$ is
the contribution for a harmonic solid.  We can then evaluate 
\begin{equation}
S_{\rm anh}(T) = \int_0^T \frac{1}{\bar{T}} \frac{\partial E_{\rm
    anh}}{\partial \bar{T}} d\bar{T}.
\end{equation}
To obtain a valid estimate of $E_{\rm anh}$ for a basin, we must heat
the IS very rapidly to insure that the system cannot change basins while
heating.  We perform such heating for at least 100 IS generated from
initial equilibrium configurations at $T=0.31$.  In principle, $E_{\rm
  anh}$ (and thus $S_{\rm anh}$) depend on the equilibrium $T$ from
which the IS are obtained; in practice, we find that the IS from
different $T$ have a nearly identical density of states $\rho(\omega)$
and $E_{\rm anh}(T)$.  Hence we can use the behavior of $E_{\rm anh}(T)$
from one set of IS for any $T$.  We find that the contribution
anharmonic contribution is rather small and negative, and can be well
described by $E_{\rm anh} = -0.065\; T$ for $T\le 0.8$; the results in
$S_{\rm anh}(T) = -0.065\; \ln T$.

\bigskip

\noindent{\large\bf \textsf {Acknowledgements}}

\bigskip

JFD acknowledges support from NIH grant 1 R01 EB006398-01A1.  FWS
acknowledges support from NSF grant number CNS-0959856 and ACS-PRF grant
51983-ND7.


\end{document}